\documentclass[letter, twocolumn, amsfonts, amssymb, amsmath, reprint, showkeys, nofootinbib, twoside,superscriptaddress]{revtex4-2}
\usepackage{graphicx,bm,hyperref,physics}
\usepackage[mathlines]{lineno}
\usepackage{xcolor}
\hypersetup{colorlinks=true,citecolor=blue,linkcolor=blue,urlcolor=black}
\begin{document}

\title{Decoding molecular distributional codes through collective instabilities}

\author{Mason N. Rouches}
\email{rouchesm@uchicago.edu}
\affiliation{James Franck Institute, University of Chicago, Chicago, IL 60637}

\author{Dongyang Li}
\affiliation{Division of Biology and Biological Engineering, California Institute of Technology, Pasadena, CA 91125}

\author{Michael B. Elowitz}
\affiliation{Division of Biology and Biological Engineering, California Institute of Technology, Pasadena, CA 91125}
\affiliation{Howard Hughes Medical Institute}

\author{Arvind Murugan}
\affiliation{James Franck Institute, University of Chicago, Chicago, IL 60637}

\begin{abstract}
Biological information is often encoded in molecular variants that differ in just a few chemical traits, such as the number of phosphorylated sites or ubiquitin chain length, rather than in arbitrarily distinct species. These molecular distributions carry information about cellular state, yet reading them with conventional molecular circuits requires prohibitively many distinct sensors. In contrast, we show that collective physical instabilities can naturally integrate the information encoded in such distributions. Using an information-theoretic matching condition between an encoded distribution and a physical readout, we derive a geometric condition that any good decoder must satisfy, and establish that phase separation, percolation, and membrane curvature instabilities all approach it for biologically natural distributions while simple mass-action binding does not. Using mean-field theory and lattice Monte Carlo simulations, we find that phase separation reads the shape of a distribution beyond its mean, robustly capturing its variance and, more weakly, its skewness, whereas mass-action binding detects only the mean. Near phase boundaries the readout captures nearly all the information present in the molecular population. Finite valency, through the threshold for network formation, adds discriminatory power invisible to mean-field theory. These results suggest that cells can exploit collective physical instabilities as natural, compact, yet near-optimal sensors for decoding molecular distributional codes.
\end{abstract}

\maketitle

Biological information is often stored in molecular identity, i.e., in which molecules are present and in what amounts. To use that information, downstream processes must discriminate among molecules well enough to route them into distinct actions. When molecular classes are cleanly separable, a small set of sensors can reliably map composition to an appropriate response.

However, cellular information is not always carried by distinct, arbitrarily different species. Instead, it is often encoded in families of closely related molecular variants that differ only along a few chemical traits. In this case, a cellular condition then corresponds to a composition profile across that trait: the abundances $P(\mathbf{n})$ of molecules with trait value $\mathbf{n}$ (e.g., the number and pattern of phosphorylations). Two environmental or cellular conditions may produce two overlapping distributions $P_1(\mathbf{n})$ and $P_2(\mathbf{n})$ that are difficult to distinguish. Such low-dimensional variation is widespread~\cite{yang2005multisite}: phosphorylation populates multiple modified forms of the same protein~\cite{roach1991multisite,wu2004selective,park2006graded}, ubiquitination generates families of chain lengths and linkage types~\cite{tsuchiya2018ub,prus2024global}, glycosylation diversifies around a limited motif set~\cite{bard2016cracking,schjoldager2020global}, many interaction domains span restricted families of binding preferences~\cite{kuriyan1997modular}, and immune cell receptors interact with diverse pools of antigen. 
Biologically relevant signals are therefore often encoded in how a population of molecules is distributed across related variants that differ along a low-dimensional trait, rather than in the identity of any single form. We will refer to such encodings as distributional codes.

How are such distributional codes read out? In principle, a chemical reaction network can extract any feature of $P(\mathbf{n})$, for instance by combining variant-specific binders or modifiers whose outputs are integrated downstream. But such circuits must be tailored to the feature of interest, with a wiring that grows with the size of the trait space. 

An alternative is to use a single collective physical process whose macroscopic response is natively sensitive to the shape of a population; such processes can couple many weak molecular interactions into a single macroscopic readout without component-by-component tabulation.  Phase separation is the most-studied example in cell biology~\cite{yoo2019cellular}, but related phenomena include polymer assembly, conformational spread and percolation in interaction networks~\cite{duke1999heightened,bray2004conformational}, and curvature-driven membrane remodeling~\cite{reynwar2007aggregation}. In each case the transition condition itself depends on the shape $P(\mathbf{n})$ of the underlying population, suggesting that collective transitions broadly are well suited to decoding distributional codes.

Here we formulate sensing as an information-theoretic matching problem between a population encoder $P(\mathbf{n};\lambda)$ and a physical decoder. Three classes of collective transitions (phase separation, percolation, and membrane remodeling) satisfy the matching condition for biologically relevant distributions, while simple mass-action binding does not. We then focus on phase separation, where biological prevalence motivates a deeper treatment. We characterize the fidelity and robustness of phase-separation-based decoding using mean-field theory and use lattice Monte Carlo simulations to show that finite valency further expands its discriminatory capacity. These results identify collective transitions as a natural physical substrate for reading distributional codes.\\~\\~\\


\begin{figure*}[htb!]
    \centering
    \includegraphics[width=\linewidth]{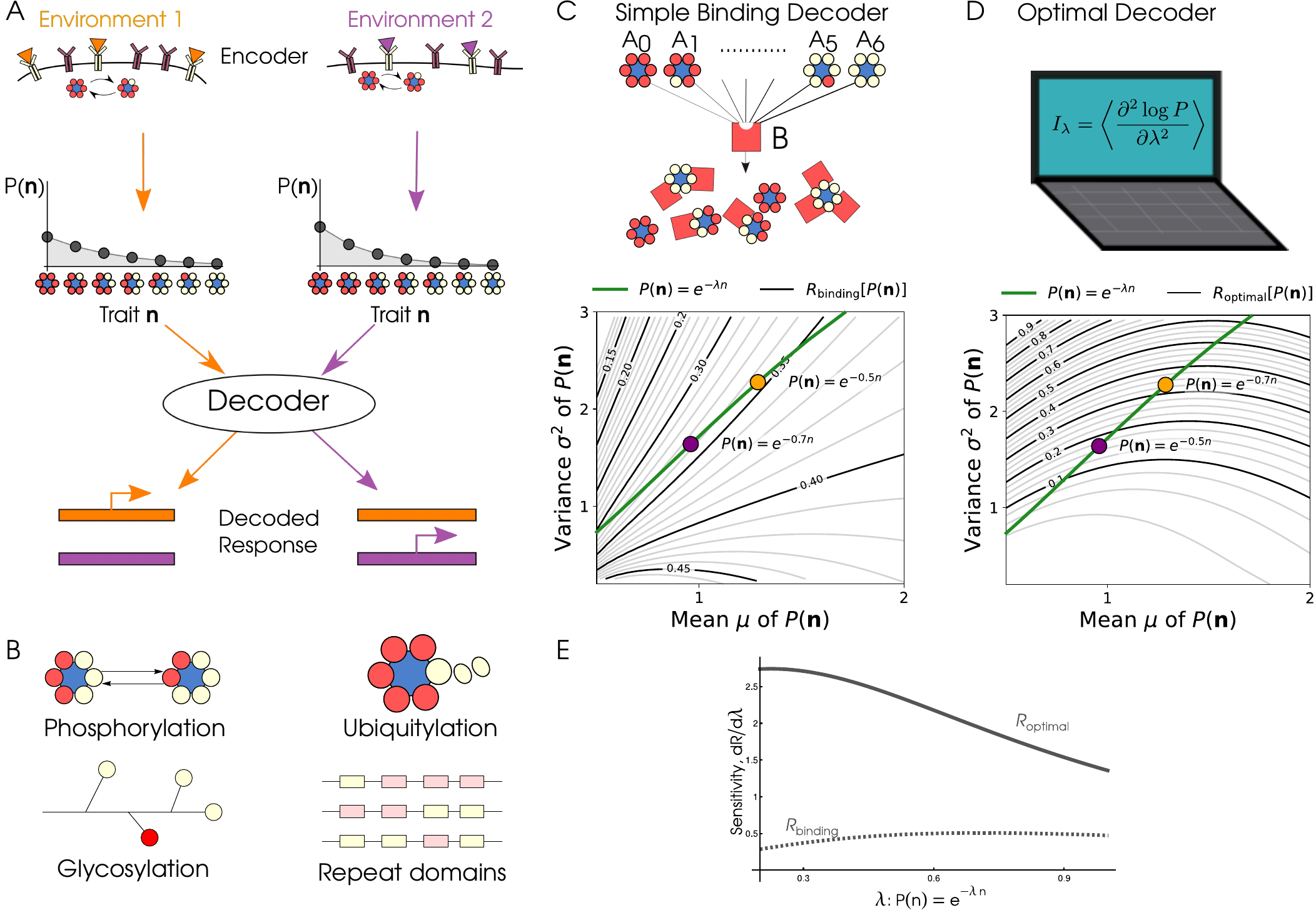}
    \caption{\textbf{Decoding molecular distributional codes} --- (A) Environmental information is encoded into a distribution $P(\mathbf{n};\lambda)$ over a molecular trait $\mathbf{n}$ by an upstream process (e.g., a signaling network), and decoded into a response, such as expression of a set of genes. A good decoder maps different environmental conditions (orange, purple) onto different responses. (B) Biological examples of low-dimensional traits: phosphorylation state, ubiquitin chain linkage, and glycosylation level or the number of repeats of a protein domain  
    (C) (top) Simple binding model: a monovalent binder $B$ interacts with molecules $A_1, A_2, \ldots, A_n$ carrying $1, 2, \ldots, n$ binding sites respectively, with abundances set by $P(n)$; the response $R_\text{binding}[P(n)]$ is the bound fraction of $A$. (bottom) An encoder that produces the exponential family $P(n)\propto e^{-\lambda n}$ traces a curve (green) in the $(\mu,\sigma^2)$ plane as $\lambda$ varies. Contours of $R_\text{binding}$ run nearly parallel to this encoder curve, so two representative distributions $P(n)$ (orange, purple dots) yield nearly the same response. (D) An optimal decoder, $R_\text{opt}[P(n)]=\sum_n P(n)e^{\alpha n}$ (SI), of the same molecules $A_n$. Contours of $R_\text{opt}$ run nearly perpendicular to the encoder curve, maximizing the difference in response between the orange and purple distributions. (E) Sensitivity $dR/d\lambda$ of response $R$ to changes in $\lambda$ of  the input distribution $P(n) \propto e^{-\lambda n}$. $R_\text{opt}$  consistently outperforms $R_\text{binding}$.
    }
    \label{fig:Fig1}
\end{figure*}

\noindent\textbf{Optimizing Information transmission by matching Encoder-Decoder pairs} ---
We formalize collective decoding as an information transmission problem. An environmental parameter $\lambda$ is encoded in a distribution $P(\mathbf{n};\lambda)$ across a low-dimensional trait $\mathbf{n}$, and a decoder maps this distribution to a physical output $R[P(\mathbf{n})]$ such as the expression of a set of target genes (Fig.~\ref{fig:Fig1}A). The Fisher information in $R$ about $\lambda$ is,

\begin{equation*}\label{eq:info}
\mathcal{I}_{\lambda} = \frac{1}{\sigma_R^2} \left( \frac{\delta R}{\delta P(\mathbf{n})} \cdot \frac{\partial P(\mathbf{n})}{\partial\lambda}\right)^2
\end{equation*}

where $\sigma_{R}^2$ is the noise in the decoder.
The integrand factorizes into a decoder response sensitivity $\delta R/\delta P(\mathbf{n})$ and an encoder direction $\partial P(\mathbf{n})/\partial\lambda$, giving a geometric matching criterion (see Supplement): contours of constant response $R$ must be orthogonal to the curve traced by $P(\mathbf{n};\lambda)$ as $\lambda$ varies. Instead, if $R$'s contours align with this curve, distinct environments collapse onto the same output and information is lost.

\begin{figure*}[htb]
\centering
    \includegraphics[width=\linewidth]{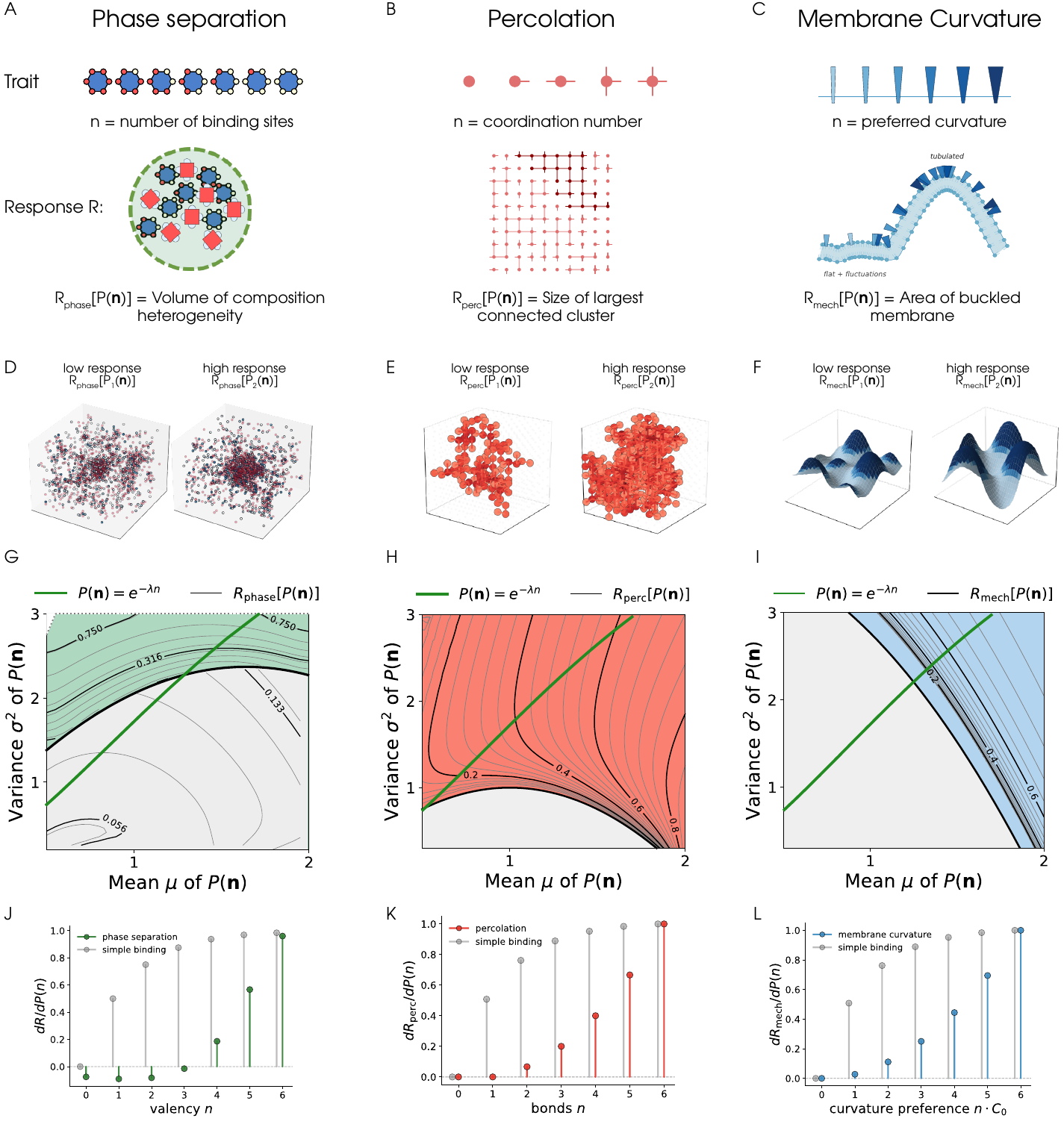}
    \caption{\textbf{Response contours for collective instabilities are naturally suited for decoding exponential distributions} --- We consider three collective physical processes, each involving molecules $A_n$ that vary along a discrete trait $n$ and with abundances $P(n)$. (A - C): Traits (upper) and responses (lower) for the three physical systems. (A)~\textbf{Phase-Separation}: The trait $n$ is the number of binding sites for multivalent sticker molecule $B$. The response $R_\text{phase}$ is compositional heterogeneity. (B)~\textbf{Percolation}: The trait $n$ is the coordination number of sites on a cubic lattice. Response $R_\text{perc}$ is the number of sites in the largest cluster. (C)~\textbf{Membrane Curvature}: The trait $n$ is the preferred curvature of a membrane protein. The response $R_\text{mech}$ is the area of membrane that buckles. (D - F): Example configurations for two distributions $P_1(\mathbf{n}),P_2(\mathbf{n})$  (from numerical simulations). In each case, $P_1(\mathbf{n})$ generates a low response, while $P_2(\mathbf{n})$ generates a large response. (G -- I) Contours of macroscopic response $R$ (gray) are plotted in the $(\mu,\sigma^2)$ plane of input distributions $P(n)$, alongside the curve (green) traced by the exponential family $P(n)\propto e^{-\lambda n}$. Response contours are nearly perpendicular to the encoded input distribution in each case. (J -- L) Sensitivity of response $R$ to components of the input distribution $P(n)$. Sensitivity curve for collective decoders (colored) is convex (as required for near-optimal decoding of exponentials) while the independent binding model curve (gray) is concave. 
    }
    \label{fig:Fig2}
\end{figure*}

\begin{figure*}[htb!]
\centering
    \includegraphics[width=\linewidth]{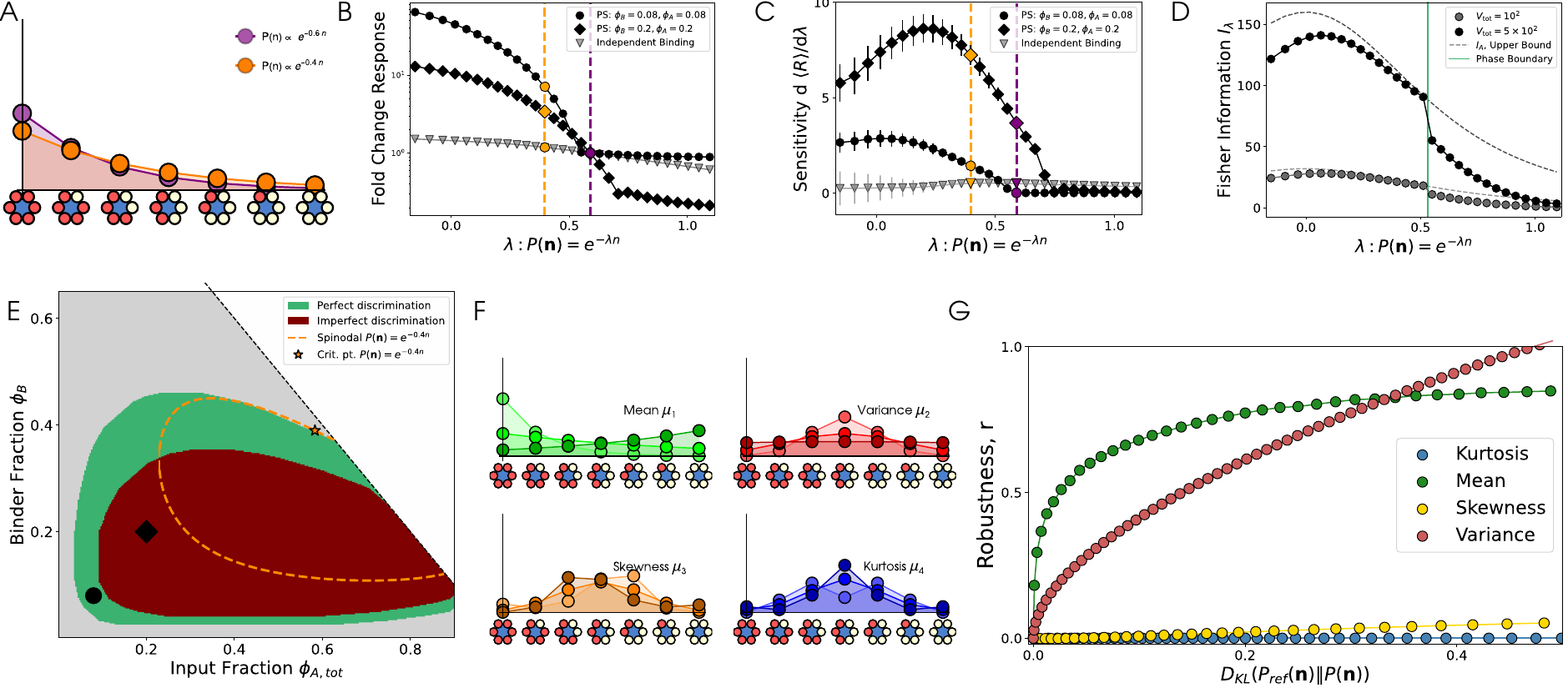}
    \caption{\textbf{Fidelity \& Robustness of Discrimination with Phase Separation} --- (A) Two input distributions $P(n)$. (B) Response $R_\text{phase}$ to distributions $P(n) = e^{-\lambda n}$ parameterized by rate $\lambda$.  Black diamonds and circles are the response of phase-separation at different system parameters. Gray triangles are the response $R_\text{bind}$ of mass action binding. Responses are normalized to the response to $P(n) = e^{-0.6n}$ (purple in A). (C) Sensitivity of the response versus distribution. Error bars denote intrinsic noise in the response, $\operatorname{Var}(R_\text{phase})$. Parameters same as in B. (D) Fisher information $\mathcal{I}_\lambda$ versus $\lambda$, for several system sizes. Near the phase boundary (green line) the information nearly saturates the total information in the system.
    (E) Discrimination regimes in hyperparameter space $(\phi_B,\phi_{A,\text{tot}})$, the volume fractions of binder $B$ and of all forms of $A$ put together. (Green) \textit{Perfect Discrimination}: only the orange input distribution in panel (A) phase-separates, distinguishable by physical state alone. (Red) \textit{Imperfect Discrimination}: both inputs in (A) phase-separate, distinguishable by droplet volume. Circular and diamond points mark the hyperparameters $(\phi_B,\phi_{A,\text{tot}})$ used in B-D. Dashed line, spinodal of $P(n)=e^{-0.4n}$; star is the critical point.
    (F) Maximum-entropy distribution families matched below moment order $i$ and varying at $i$ and above. Fixed moments $\mu_1=3$, $\mu_2=1.6$, $\mu_3=0$. 
    (G) Hyperparameter space robustness $r = \text{Perfect}/(\text{Perfect}+\text{Imperfect})$ (fractional area of green region in (E)) versus distributional distance $D_{KL}(P_\text{ref}\Vert P)$, for the families in F.
    }
    \label{fig:Fig3}
\end{figure*}

We make these ideas concrete with a one-parameter family of encoders: exponential distributions $P(n) \propto e^{-\lambda n}$ over a discrete trait $n$ such as phosphorylation number, ubiquitin chain length, or binding-domain repeat count (Fig.~\ref{fig:Fig1}B). Exponentials are both mechanistically natural and biologically widespread. They arise whenever each step of a sequential modification reaction has a roughly constant per-step efficiency, so that the abundance of $n$-fold modified species falls geometrically with $n$.  Such profiles are observed in polyubiquitin chain length distributions, sequential multisite phosphorylation, and template-free polymer/oligomer length distributions~\cite{burlacu1992distribution,romberg2001polymerization,salazar2009multisite,tsuchiya2018ub} (an exponential distribution is also the maximum-entropy distribution over $n$ given only its mean, making it the generic steady-state form when no further structure is imposed). We use this family as an illustrative case for Figs.~\ref{fig:Fig1}--\ref{fig:Fig2}; in Fig.~\ref{fig:Fig3} we relax it and ask how phase separation discriminates more general $P(n)$, including families matched in mean and variance that differ only in higher moments such as skewness and kurtosis.

We first consider a minimal decoder: a binder molecule $B$ that interacts independently with sites on the input molecules $A_n$ (Fig.~\ref{fig:Fig1}C); the response $R_\text{binding}$ is the fraction of $A$ molecules bound at least one $B$ molecule. 
We plot contours of $R_\text{binding}$ in the $(\mu,\sigma^2)$ plane of the input distribution. For the exponential family $P(n)\propto e^{-\lambda n}$, the mean and variance satisfy $\sigma^2 = \mu(\mu+1)$, so the encoder traces a near-parabola in the $(\mu,\sigma^2)$ plane that runs nearly tangent to the binding contours. By our criterion, simple binding is a poor decoder of exponential distributions: changes in $\lambda$ move the encoded distribution along, rather than across, the decoder's contours.

From our geometric picture, good decoders of the exponential family should have response contours where variance \textit{decreases} with the mean, Figure ~\ref{fig:Fig1}D. As shown in the SI, such good decoders must weight high-$n$ species more than linearly with $n$ and in fact, the optimal decoder has exponential weighting, i.e., $R_\text{opt}[P(n)] = \sum_n P(n) e^{\alpha n}$ (Fig.~\ref{fig:Fig1}D). Comparing sensitivities $dR/d\lambda$ across the exponential family confirms that $R_\text{opt}$ outperforms simple binding for all $\lambda$, with the gap widening as the encoder curve becomes more parallel to the binding contours (Fig.~\ref{fig:Fig1}E).

\noindent\textbf{Collective Instabilities decode exponential distributions} --- What physical systems naturally approach the performance of the optimal decoder?  

We extend the binding model by letting the binder $B$ be multivalent (Fig.~\ref{fig:Fig2}A). The encoded input is still a distribution $P(n)$ of molecules $A_n$ carrying $n$ sticker sites, but each binder $B$ now exposes multiple binding sites and can thus bridge several $A_n$'s. This system is described by the physics of phase separation with a free energy:

\begin{align*}
\frac{F}{Vk_{B}T} &= \sum_{n=0}^{N_s} \phi_{A,n}\log\phi_{A,n} + \frac{n J_\text{B}}{k_{B}T}\,\phi_{A,n}\phi_\text{B}\\
&+\phi_\text{B}\log{\phi_\text{B}} + \left(1-\phi_B-\phi_{A,\text{tot}}\right)\log{\left(1-\phi_B-\phi_{A,\text{tot}}\right)}
\end{align*}

where $\phi_{A,n}$ is the volume fraction of the encoded input molecule $A_n$, $\phi_{B}$ is the volume fraction of the multivalent binder $B$, and $J_\text{B}$ is the interaction energy between a sticker site on $A_n$ and any site on $B$. The equilibrium state of the system is determined by minima of $F$. We define the response $R_\text{phase}[P(n)]$ as the volume of the dense phase enriched in $A$ and $B$ molecules, if it exists. When phases do not coexist, we define $R_\text{phase}[P(n)]$ as the correlation volume, or typical length scale of compositional fluctuations.

In Figure~\ref{fig:Fig2}G we plot response contours of the phase-separating system in the plane of distributions $P(n)$ parameterized by mean $\mu$ and variance $\sigma^2$. The phase-separation response $R_\text{phase}[P(n)]$ cuts across the encoding contour for exponential distributions $P(n) \propto e^{-\lambda n}$ nearly orthogonally. 

Another broadly relevant collective physical process that is sensitive to a low-dimensional trait distribution is percolation, Figure ~\ref{fig:Fig2}B. Here the trait value sets the number $(0,\dots, 6)$ of available bonds per site on a cubic lattice, and the response $R_\text{perc}[P(n)]$ is taken to be the size of the largest connected cluster. The probability of adding a bond to a cluster depends only on the mean of the bond distribution, $p_\text{bond} \propto \mu$, while the statistics of cluster sizes depend on the entire distribution (see SI). For instance, the percolation threshold traces the curve $\sigma^2 = 2\mu-\mu^2$ in the space of distributions. Mean-field theory and Monte Carlo simulations show that contours of $R_\text{perc}[P(n)]$ are aligned to exponential distributions in the $\mu-\sigma^2$ plane, Figure ~\ref{fig:Fig2}H. The decoder's sensitivity to the distribution, $\delta R_\text{perc}[P(n)]/\delta P(n)$ is convex in the bond number $n$, Fig.~\ref{fig:Fig2}K, confirming that percolation is matched to exponential distributions.

Mechanical systems are also collective decoders. We consider an example of membrane bending by a family of curvature-generating proteins. Here we take a series of individual proteins to favor spontaneous membrane curvatures between $0$ (flat) and $6 c_0$ (highly curved), where $c_0$ is an intrinsic preference, Figure ~\ref{fig:Fig2}C. Each protein locally bends the membrane towards its preferred curvature. Crucially, highly curved regions preferentially recruit proteins favoring that curvature, leading to inhomogeneous curvature throughout the membrane -- some regions might go unstable and buckle while others remain flat. The resulting equilibrium state of the membrane balances these curvature preferences, protein mixing entropy $f_\text{mix}(\vec{\phi})$, and tension-associated costs $\sigma_\text{mem}$~\cite{leibler1986curvature} (see Supplement for our detailed model). The free energy $F_\text{mem}$ of a membrane with mean curvature $C$ and protein distribution $\vec{\phi}$  is

\begin{equation}
F_\text{mem}[C, \vec{\phi}] = \int dS \left[ \sigma_\text{mem}+\frac{\kappa}{2}\bigl(C-C_0(\vec{\phi}))^2+ f_{\text{mix}}(\vec{\phi}) \right]
\end{equation}

where $C_0(\vec{\phi}) \equiv c_{0,\text{mem}}+ \sum_n c_0 n\, \phi_n(\mathbf{r})$ is the protein-modified spontaneous curvature preference, and $\kappa$ the curvature modulus. We define the response $R_\text{mech}[P(n)]$ to be the fraction of buckled membrane, which is determined by minimization of $F_\text{mem}$ over protein densities and membrane shape (see Supplement).  Again, these contours are nearly orthogonal in the $\mu-\sigma^2$ plane, Fig.~\ref{fig:Fig2}I, and the decoder's sensitivity is convex in the trait value, Fig. ~\ref{fig:Fig2}L.

All three phenomena here exploit \textit{collective instabilities} that naturally couple to the full shape of the trait distribution $P(\mathbf{n})$, not merely its mean. While a bespoke molecular circuit can be constructed to sense each trait value $P(\mathbf{n})$ individually and integrate them in just the right way, these collective instabilities perform that integration for free, yielding response contours that are nearly optimal in an information-theoretic sense. \\

\noindent\textbf{Fidelity and Robustness of a phase-separation based decoder} --- The appeal of collective decoding is its simplicity but it requires the system to sit near a collective instability. Consider two distinct environments encoded in the distributions of Fig. ~\ref{fig:Fig3}A. To distinguish these two distributions, the decoder's response to one distribution must be much larger than it is for the other. For some system parameters, 
phase-separation based decoders perform well, while for others this amplification is poor, Fig. ~\ref{fig:Fig3}B. While evolution can plausibly tune parameters to keep a biological system near such an instability, this mode of sensing is broadly relevant only if the instability is accessible across a wide region of parameter space, not just at a fine-tuned point. 

We now ask how precisely phase-separation based decoding can distinguish two input distributions $P(\mathbf{n})$ at fixed hyperparameters given intrinsic noise (\textit{fidelity}), and how large a region of hyperparameter space supports such discrimination (\textit{robustness}). The relevant hyperparameters are the interaction strength $J_{B}$ and the binder and input volume fractions $\phi_B,\phi_{A,\text{tot}}$. We first hold these hyperparameters fixed and characterize fidelity, then sweep over them to characterize robustness.

At a fixed hyperparameter choice, the intrinsic noise in the response $R$ sets a precision floor on estimating $\lambda$, captured by the Fisher information $I_\text{phase}$. Combined with the noise floor on the input molecules, $I_\text{A} = \sigma^2\phi_{A,\text{tot}}V_\text{tot}$, the Fisher information on $\lambda$ is $\mathcal{I}_\lambda = \frac{I_A I_\text{phase}}{I_A + I_\text{phase}}$. As $\lambda$ varies, the system passes through three regimes with different information content (Fig.~\ref{fig:Fig3}C,D).

When input distributions do not drive phase separation, the response is small and the noise is on the same scale Fig.~\ref{fig:Fig3}C. Here the information captured by phase-separation is small and a fraction of the total information available Fig.~\ref{fig:Fig3}D. At values that do drive phase separation, the response is large with moderate noise, and phase-separation captures a large fraction of the available information. Near the threshold $\lambda$ at which the system phase-separates, the response is large and the noise is small. Here phase-separation conveys nearly all of the available information, Fig~\ref{fig:Fig3}D (green line). 

No decoder can exceed the information limit set by counting individual molecules. What phase-separation can do is leverage the information present in larger systems by aggregating more molecules into a collective measurement: in most of parameter space the Fisher Information scales with system size, $I_\text{phase} \sim V_\text{tot}$. Near critical points and first-order phase-boundaries, however, this scaling is greater than linear (see Supplement). Since $I_\text{phase}$ increases with system size, phase-separation can integrate most of the available information into a collective measurement, and near a phase-boundary can approach the limit set by molecular counting. 

The Cramér-Rao bound,  $\operatorname{Var}(\delta\lambda) \ge 1/\mathcal{I}_\lambda$, implies that $\mathcal{I}_\lambda \approx 10^2$ can constrain an estimate of $\lambda \pm 0.1$, sufficient to discriminate the two distributions in Fig.~\ref{fig:Fig3}A (separated by $\lambda_1-\lambda_2 \approx 0.2$), but only near the phase boundary.

Turning to robustness, we now ask how large a region of hyperparameter space supports reasonable fidelity at all. Overlaying the equilibrium phase diagrams of the two input distributions in the $(\phi_B,\phi_{A,\text{tot}})$ plane (Fig.~\ref{fig:Fig3}E) defines three regions. Where only one distribution phase-separates, the two are distinguishable by physical state alone and amplification can be made arbitrarily large; we call this \textit{Perfect Discrimination}. Where both phase-separate, discrimination requires comparing droplet volumes and amplification saturates at the system scale; we call this \textit{Imperfect Discrimination}. Outside both, only weak dilute-phase amplification is available, except near critical points. For the two example distributions in Fig.~\ref{fig:Fig3}A, the Perfect and Imperfect zones together occupy a substantial fraction of the $(\phi_B,\phi_{A,\text{tot}})$ plane. Discrimination is therefore accessible across a broad swath of hyperparameter space rather than only at a fine-tuned point.

Next we quantify the size of these zones in hyperparameter space $(\phi_B,\phi_{A,\text{tot}})$ plane as a measure of robustness. We define robustness by $r = \text{Perfect}/(\text{Perfect}+\text{Imperfect})$, plotted against the Kullback-Leibler divergence $D_{KL}(P(n) \Vert Q(n))$ between the two distributions (Fig.~\ref{fig:Fig3}G). For distributions that differ only by mean or only by variance, the robustness $r$ increases quickly as the input distributions become more distinguishable (i.e., higher $D_{KL}(P(n)\Vert Q(n))$). Hence, our mean-field theory description of phase separation predicts robust discrimination of input distributions over a large region of hyperparameters.

We then asked whether phase separation can discriminate distributions that share the same mean and variance but differ in higher-order moments such as skewness or kurtosis. To pose these tasks, we generated families of maximum-entropy distributions with matched lower moments and varying higher ones (Fig.~\ref{fig:Fig3}F). The robustness curves for skewness and kurtosis saturate at very low values (Fig.~\ref{fig:Fig3}G), indicating that these tasks remain difficult for phase separation even when the distributions are highly distinguishable. This asymmetry between low and high moments has an analytical origin in the structure of the phase boundary. The spinodal condition (Supplement),
\begin{equation}\label{spinodalNew}
\frac{\phi_B\left(1-\phi_B\right)\left(\sigma^2+\mu^2\left(1-\phi_{A,\text{tot}}\right)\right)\phi_{A,\text{tot}}}{\left(\phi_{A,\text{tot}}\mu\phi_B + k_B T/J_\text{B}\right)^2} = 1,
\end{equation}
depends on $\mu$ and $\sigma^2$ but not on higher moments of $P(n)$, while critical-point conditions add a dependence on the skewness $\mu_3$. Phase diagrams of distributions that differ at the level of kurtosis thus have identical critical points and spinodals, limiting a large region of parameter space to imperfect discrimination. This implies a vanishing bound on robustness for kurtosis and higher moments, and a non-zero bound for mean, variance, and skewness discrimination. 

These results treat binders and inputs as independent. Finite valency can saturate available binding sites in ways that may expand or constrain the discriminable region, which we address next with lattice Monte-Carlo simulations.\\

\begin{figure*}[htb!]
\centering
    \includegraphics[width=\linewidth]{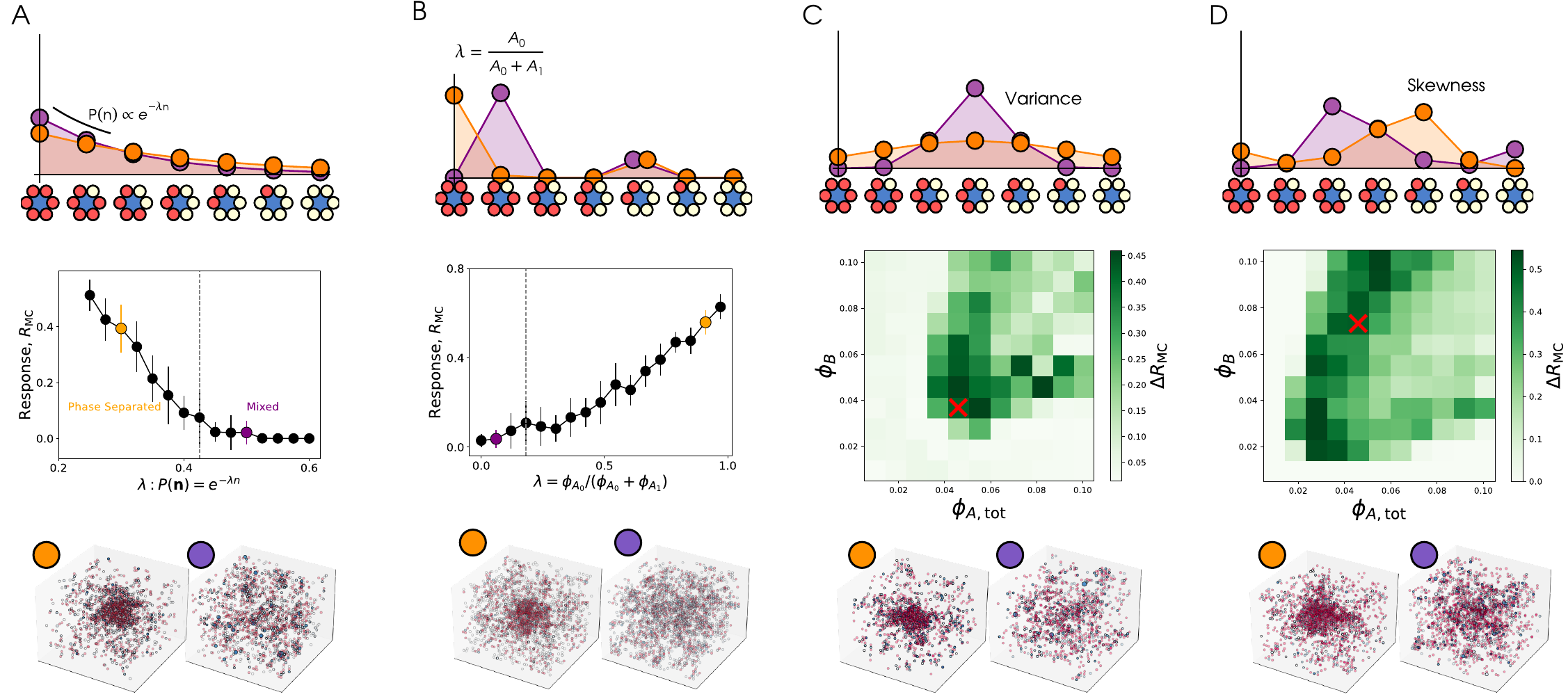}
    \caption{
    \textbf{Finite valency effects, captured through Monte-Carlo simulations, enhance discrimination} ---
    (A) We used Monte-Carlo simulations to discriminate the (i) two exponential distributions shown. (ii) Response $R_\text{MC}[P(\mathbf{n})]$ is the output of a linear classifier on the binned cluster size distribution.  Dashed line is the estimated phase-boundary. (iii) Representative configurations from simulations of the two distributions. 
    (B) Same analysis as in (A) but to discriminate distributions with the same multivalent content ($\phi_{A_4}$) and differ entirely in the amount of singly-valent complexes $\lambda = \frac{\phi_{A_{0}}}{\phi_{A_1}+\phi_{A_0}}$. Distributions with high amounts of singly-valent molecules sequester binders, inhibiting phase-separation. 
    (C) Robustness analysis for (i) input distributions with equal mean binding sites and different variance. (ii) Discrimination between the two distributions as a function of hyperparameters $\phi_B, \phi_{A, \text{tot}}$. Color is a proxy for fidelity and indicates the difference in response between the two distributions, $\Delta R_\text{MC} = R_\text{MC}[P_1(\mathbf{n})] - R_\text{MC}[P_2(\mathbf{n})]$. Darker points are more distinguishable than lighter points.
    (D) Same analysis as in (C) for input distributions that differ in skewness alone but equal mean and variance. 
    }
    \label{fig:Fig4}
\end{figure*}

\noindent\textbf{Discriminating molecular distributions with Monte-Carlo simulations} ---
Mean-field theory weights each molecule $A_n$'s contribution to phase separation by its sticker count $n$, but misses a key aspect of phase separation: molecules with $n \le 2$ cannot drive phase separation since network assembly requires $n \gtrsim 3$~\cite{semenov1998thermoreversible,choi2020physical}. We therefore expect finite valency to \textit{enhance} discriminatory capacity beyond the mean-field bounds of Fig.~\ref{fig:Fig3}G, because, e.g., distributions $P(\mathbf{n})$ matched in mean but differing in mass above the bridging threshold $n = 3$ should give qualitatively different responses.

To test this, we performed lattice Monte-Carlo simulations of an explicit-valency analogue of the Fig.~\ref{fig:Fig2}A model. Each input molecule $A_n$ carries exactly $n$ binder-binding sites with $n \in \{0,\dots,6\}$ and each binder $B$ carries $n_B=4$ sites; particles occupy a cubic lattice and exchange under particle-swap and bond-rearrangement moves accepted with a Metropolis criterion (see Supplement for details). The readout $R_\text{MC}[P(\mathbf{n})]$ is a linear classifier on the binned cluster-size distribution, with weights fit to separate the two target distributions.

For exponential inputs, $P(n)\propto e^{-\lambda n}$,    at fixed hyperparameters, the readout switches sharply with $\lambda$ across the simulated phase boundary (Fig.~\ref{fig:Fig4}A), reproducing the mean-field picture of Fig.~\ref{fig:Fig1}D. The bridging threshold then produces a discrimination axis invisible to mean-field theory. Mixing $A_0$, $A_1$, and $A_4$ at fixed multivalent content $\phi_{A_4}$ while replacing $A_0$ with $A_1$ raises the mean sticker count, which should promote phase separation in the mean-field picture; in simulation it does the opposite (Fig.~\ref{fig:Fig4}B), because each $A_1$ caps a site on the binder without contributing to bridging. The same trait shift moves the mean-field and lattice responses in opposite directions. Finally, we find that discrimination of higher moments is more robust than mean-field predicts. To quantify this we mapped the response $R_\text{MC}$ across hyperparameters $(\phi_B,\phi_{A,\text{tot}})$ for distributions that differ in variance (Fig. ~\ref{fig:Fig4}C) and skewness (Fig.~\ref{fig:Fig4}D). Both pairs admit broad discriminable regions of high fidelity where the differential response, $\Delta R_\text{MC} =R_\text{MC}(P_1) - R_\text{MC}(P_2)$, is large. Skewness is not visibly harder than variance, partially lifting the mean-field bound of Fig.~\ref{fig:Fig3}G.\\

\noindent\textbf{Discussion} ---

Our central result is that collective physical instabilities are natural decoders of molecular distributional information. Information theory predicts a geometric matching condition, namely that efficient decoders must have response contours crossing the encoder trajectory orthogonally. Phase separation, percolation, and membrane curvature remodeling all satisfy this condition for the encoder families studied here, while simple mass-action binding does not (Figs.~\ref{fig:Fig1}--\ref{fig:Fig2}). For phase separation specifically, this sensitivity can be understood analytically; the mean-field spinodal (Eq.~\ref{spinodalNew}) is already sensitive to $\mu$ and $\sigma^2$, even before accounting for the basic network-formation requirement of $n \geq 3$ binding sites needed for phase separation. Lattice Monte Carlo simulations incorporating finite valency expand this further, yielding sensitivity to $\mu_3$. Both model classes are equilibrium; non-equilibrium drive remains an important open extension.

Compared to simple binding networks, such instability-based decoding is more compact in that it collapses sensing and response into a single physical act. For example, consider how proteins marked for destruction through ubiquitination are sensed and destroyed (by autophagy or the proteasome)~\cite{lu2017receptor,sun2018polyubiquitin}. Ubiquitin ligases polymerize chains of ubiquitin on proteins, earmarking them for degradation. Polyubiquitinated proteins promote the condensation of specific (autophagocytic) degradatory machinery, while monoubiquitinated proteins do not~\cite{sun2018polyubiquitin,fujioka2020phase}. In this way sensing (condensation) and actuation (substrate selection and degradation) are a concerted physical process.

Our results extend prior work in two directions. Studies of valency effects on phase boundaries have focused on monodisperse systems~\cite{nandi2022affinity}; we show that the distribution of valencies can qualitatively reshape phase diagrams. The broader multicomponent phase-separation literature has largely considered molecules with statistically random and distinct interactions~\cite{sear2003instabilities,shrinivas2021phase,zwicker2022evolved}, modeling a regime where individual species are highly dissimilar. We study the opposite limit, of molecules that are nearly identical, differing only along a low-dimensional trait. Polydispersity theory~\cite{sollich2001moment,graf2022thermodynamic} established that phase behavior depends on the full composition of a mixture; we add an information-theoretic lens that asks which distributional features are discriminable and at what cost in decoder parameter tuning.

Distributional codes are inexpensive to generate. Template-free modification processes (polyubiquitination, multisite phosphorylation) produce statistical profiles across molecular variants without requiring enzymatic specificity for each form~\cite{strickfaden2007mechanism,mukhopadhyay2016multisite}, and the resulting signal is robust since it is a statistical property of a population rather than the precise state of any one molecule. Structured sequential modification processes such as kinetic proofreading also generate exponential distributions over modification states as a natural byproduct, ~\cite{strickfaden2007mechanism,kirby2023proofreading}, placing these encoders precisely in the class for which collective instabilities serve as natural decoders. 

A concrete example occurs in the discrimination of self from non-self ligand by the T-cell receptor (TCR).  The statistics of ligand (un)binding events drive a concert of phosphorylations that culminates in the multi-site phosphorylation of order hundred transmembrane LAT molecules~\cite{lo2019slow,mcaffee2022discrete}. Phosphorylated LAT phase-separates on the plasma membrane to form large signaling complexes~\cite{su2016phase}, that polymerize actin and can lead to T-cell activation. Here the TCR encodes binding statistics into a distribution over LAT phosphorylation states. This distribution is decoded by phase-separation, which combines the noisy measurement and weak interactions in every LAT molecule into an informative, macroscopic readout. 

In B-cell affinity maturation, collective physical readouts of antigen affinity can be more informative than molecular level readouts~\cite{knevzevic2018active,jiang2024physical}. B-cells are selected on the basis of their ability to bind to specific antigen, though activation itself could be sensitive to antigen distributions.  

Our work suggests several natural extensions. Cooperative binding, internal linker flexibility~\cite{xu2020rigidity,linne2024optimality}, and multi-component condensates could expand discriminatory capacity, and the encoder-decoder matching framework provides a principled criterion for evaluating each. Phase morphologies may further serve as readouts for sequential reactions at condensate interfaces~\cite{quinodoz2025mapping}. More broadly, these results suggest a rationale for why biological information is so often carried by distributions over related molecular variants rather than by arbitrarily distinct species. Both the encoding (imprecise modification) and the decoding (collective assembly near a thermodynamic instability) exploit processes that are physically natural at the molecular scale, rather than demanding fine-tuned specificity at either stage.

\noindent\textit{Acknowledgments} --- We thank Isabella Graf, Max Schelling, Pepijn Moerman, Henry Alston, Krishna Shrinivas, Mats von Tongeren, Samantha Stam, and members of the Murugan group for useful discussions. This work was supported by the National Science Foundation through the Center for Living Systems (grant no. 2317138). A.M. acknowledges support from the NIGMS of the National Institutes of Health under award number through R35GM151211. M.B.E. is a Howard Hughes Medical Institute investigator, and was also supported by the Alfred P. Sloan Foundation (award number G-2024-22436) and the Chan Zuckerberg Initiative (award number 2024-349887)
\bibliography{refs}

\pagebreak

\end{document}


\title{Supplementary Material for Discriminating Molecular Distributions with Phase-Separation}
\author{Mason N. Rouches}
\email{rouchesm@uchicago.edu}
\affiliation{James Franck Institute, University of Chicago, Chicago, IL 60637}

\author{Dongyang Li}
\affiliation{Division of Biology and Biological Engineering, California Institute of Technology, Pasadena, CA 91125}

\author{Michael B. Elowitz}
\affiliation{Division of Biology and Biological Engineering, California Institute of Technology, Pasadena, CA 91125}
\affiliation{Howard Hughes Medical Institute}

\author{Arvind Murugan}
\affiliation{James Franck Institute, University of Chicago, Chicago, IL 60637}
\maketitle

\section*{Outline of Supplement}
In Section ~\ref{sec:info} we define the general encoder-decoder framework. We show how encoder-decoder matching emerges from optimizing information transmission, and demonstrate what mathematical properties physical encoders and decoders must satisfy in order to match. In Section ~\ref{sec:binding} we detail the 'simple binding' model that appears in the main text, and calculate quantities relevant for information transmission. In Section ~\ref{sec:phase} we present an in-depth analysis of the phase-separation model we utilize in the main text. First we analyze the phase-behavior, locating regions of instability, critical points, and phase-coexistence. We then present the information theoretic analysis of phase-separation as a decoder. In Section ~\ref{sec:mc} we supply our model for Lattice Monte-Carlo simulations of phase-separation. In section ~\ref{sec:perc} we present mean-field theory and numerical simulations for the percolation model explored in the main text. In section ~\ref{sec:membrane} we present the membrane curvature model presented in the main text, substantiated with numerics.\\

\section{Optimizing Information Transmission with Encoder-Decoder Matching}\label{sec:info}
\noindent\textit{Information Content of a Distribution} -- We consider systems where information about an environmental parameter $\lambda$ is spread across molecular species whose abundances are described by a distribution $P(\mathbf{n};\lambda)$. An \textit{Encoder} maps the environment onto a distribution. A \textit{Decoder} $R[P(\mathbf{n})]$ maps this distribution into a single, scalar response. The sensitivity of the decoder to the environment, $\frac{\partial R}{\partial\lambda}$, factorizes into $\frac{\partial R}{\partial\lambda} = \frac{\delta R}{\delta P(\mathbf{n})} \cdot \frac{\partial P(\mathbf{n})}{\partial\lambda}$ -- the change in readout is an inner product of the decoder's variation with the distribution and the encoder's variation with the environment.  Inferring a change in $\lambda$ from the value of $R[P(\mathbf{n})]$ will be limited by the Fisher information $\mathcal{I}_\lambda$: 
\begin{align*}
\mathcal{I}_{\lambda} &= \int  \left(\frac{\partial \log P(R;\lambda)}{\partial \lambda} \right)^2P(R;\lambda) dR\\
& = \int \left(\frac{\partial P(R;\lambda)}{\partial\lambda}\right)^2 \frac{1}{P(R;\lambda)}dR\\
&\approx\left(\frac{1}{2}\left(\frac{\partial \langle R \rangle}{\partial\lambda} \right)^2 + \left(\frac{\partial\sigma_R^2}{\partial\lambda}\right)^2\right) \frac{2}{\sigma_{R}^2}\\   
&=  \left(\frac{\delta \langle R\rangle}{\delta P(\mathbf{n})} \cdot \frac{\partial P(\mathbf{n})}{\partial\lambda}\right)^2 \frac{1}{\sigma_\text{dec}^2+\sigma_\text{enc}^2}+
\frac{\left(\partial_{\lambda}\sigma^2_\text{dec}+\partial_{\lambda}\sigma^2_\text{enc}\right)^2}{(\sigma_\text{dec}^2+\sigma_\text{enc}^2)^2}
\end{align*}

where $P(R;\lambda)$ is the probability distribution of responses at a particular $\lambda$, $\langle R\rangle,\quad\sigma^2_{R} = \sigma^2_\text{dec} + \sigma^2_\text{enc}$ are the mean and variance of the readout distribution $P(R;\lambda)$. We have assumed $R$ is Gaussian distributed to go from line two to line three, and that $\sigma^2_{R}$ is independent of $\lambda$. We have decomposed the total readout noise into two independent contributions. The decoder noise $\sigma^2_\text{dec} = \operatorname{Var}(R)\vert_{P(R;\lambda)}$ captures the fluctuations in the readout for a fixed distribution, and $\sigma^2_\text{enc} = \operatorname{Var}_{P(\mathbf{n})}(\langle R\rangle)$ is the fluctuation in the mean readout from distribution noise. In the main text and throughout, we neglect the final term, assuming that the encoder and decoder noise vary slowly with the environment. Together this brings us to the Fisher information presented in the main text:
\begin{equation}\label{eq:finfo}
\mathcal{I}_{\lambda}=\left(\frac{\delta \langle R\rangle}{\delta P(\mathbf{n})} \cdot \frac{\partial P(\mathbf{n})}{\partial\lambda}\right)^2 \frac{1}{\sigma_\text{dec}^2+\sigma_\text{enc}^2}
\end{equation}

\noindent\textit{Gain} --- To arrive at our geometric picture of information transmission, we separately analyze the gain (numerator), and both noise components (denominator) of Equation ~\ref{eq:finfo}. The gain term is split into decoder and encoder contributions. Decomposing the decoder into an amplitude and direction,  $\frac{\delta R}{\delta P(n)} = A_\text{R}\vec{u}_\text{R}$, separates the per-component changes $\vec{u}$ from the `raw-gain'  $A_\text{R}$. The encoder's variation with the environment can be similarly decomposed, $\frac{\partial P}{\partial\lambda} = A_\text{P}\vec u_\text{p}$; the dot product can then be written as
\begin{align*}
\left(\frac{\delta \langle R\rangle}{\delta P(\mathbf{n})} \cdot \frac{\partial P(\mathbf{n})}{\partial\lambda}\right)^2 &=  \left(A_P A_R \vec{u}_P\cdot\vec{u}_{R}\right)^2
\end{align*}
This expression relates the total gain to encoder and decoder specific amplitudes $A_\text{P},A_\text{R}$, and orientation $\vec{u}_P\cdot\vec{u}_R$. Optimizing a decoder constitutes choosing the $\vec{u}_R$ to maximize $\mathcal{I}_\lambda$; in a noiseless setting this means matching the two directions, $\vec{u}_R\cdot\vec{u}_P = 1$ \\

\noindent\textit{Decoder Noise} --- Decoder noise depends on system specific details, which we compute for phase-separation in Section ~\ref{sec:phase}. Here we analyze generic extensive, thermodynamic fluctuations in the decoder. We assume that the strength of noise scales with the readout itself, $\sigma_\text{dec} \propto \langle R\rangle^{1/2}$, such that the noise effectively vanishes in the thermodynamic limit: 
\begin{align*}
\sigma_\text{dec}^2 &= \langle R\rangle\sigma_0^2
\end{align*}
where $\sigma_0$ is the scale of the decoder noise, and can be calculated once a system is specified.
\\
\noindent\textit{Encoder Noise} --- To compute encoder noise, we push the response kernel through the intrinsic variance in $P(\mathbf{n})$ given shot noise from limited particle numbers, and noise in the encoding process itself. Shot noise is specified by the covariance matrix $\Sigma_\text{shot} \equiv N_\text{tot}^{-1}\Sigma_{P} = N_\text{tot}^{-1}\left(\operatorname{diag}(\vec{P}) - \vec{P}\vec{P}^\top\right)$. Encoder noise is system-specific and captured by a covariance matrix, $\Sigma_\text{enc}$. Below we compute the noise on readout arising from these two sources:
\begin{align*}
\sigma_\text{enc}^2 &= \left(\frac{\delta R}{\delta P(n)}\right)\left(\Sigma_\text{shot} +\Sigma_\text{enc}\right)\left(\frac{\delta R}{\delta P(n)}\right)\\
&=A_\text{R}^2\left(\vec{u_\text{R}}^T\left(N_\text{tot}^{-1}\Sigma_\text{P} + \Sigma_\text{enc}\right)\vec{u_\text{R}}\right)\\
&=\frac{A_\text{R}^2}{N_\text{tot}}\underbrace{\operatorname{Var}_{P(n)}(\vec{u}_\text{R})}_{\sigma^2_\text{shot}} \\&+ A_\text{R}^2\underbrace{\left(\left(A_Pa_\parallel\vec{u}_P\cdot\vec{u}_R\right)^2 + \left(a_\perp\vec{u}_\perp\cdot\vec{u}_R\right)^2+2a_\times\left(\vec{u}_P\cdot\vec{u}_R\right)\left(\vec{u}_\perp\cdot\vec{u}_R\right)\right)}_{\sigma^2_\text{enc}}
\end{align*}
Here we have decomposed the encoder covariance into terms components that are `parallel' to the signal $a_\parallel$, those that are perpendicular $a_\perp$, and cross terms with the $a_{\times}$. \\

\noindent\textit{Optimizing Information Transmission} --- Putting these pieces together, we compute the Fisher Information on $\lambda$ :
\begin{align*}
\mathcal{I}_{\lambda} &= \frac{\left(A_{P}A_{R}\vec{u}_R\cdot \vec{u}_p\right)^2}{\sigma_{0}^2\langle R \rangle + A_{R}^2\left(N_\text{tot}^{-1}\sigma^2_\text{shot}  + \sigma^2_\text{enc} \right)}\\
&=\frac{\left(A_{P}\vec{u}_R\cdot \vec{u}_p\right)^2}{\left(\vec{u}_R\cdot \mathbf{M} \vec{u}_R\right)}\\
\end{align*}
with the noise defined via the matrix $\mathbf{M}$:
\begin{equation*}
\mathbf{M}_{i,j} = \delta_{i,j}\frac{\sigma_0^2\langle R\rangle}{A_{R}^2} + \frac{\delta_{i,j}P_i -P_iP_j}{N_\text{total}}  + \Sigma_{i,j,\text{enc}}
\end{equation*}

To find the optimal decoder, we have to choose the best direction for $\vec{u}_{R}$. In order to buttress noise and remain signal - aligned, the optimal decoder must be have $\vec{u}_{R}\propto \mathbf{M}^{-1}\vec{u_p}$. If $\mathbf{M}^{-1}$ has uniform entries, the previous intuition of maximizing $\vec{u}_R\cdot\vec{u}_P$ holds. If not, the optimal readout will be rescaled to match the statistics of the noise. 

Plugging this in we get the optimized Fisher Information:
\begin{align*}
\mathcal{I}_{\lambda}^{\star} 
&=A_p^2 \vec{u}_p \mathbf{M}^{-1}\vec{u}_p
\end{align*}

To find the relevant weights we analyze the above expression in several limits. First, to investigate the scaling behavior, we assume that the decoder is an extensive, thermodynamic system: $A_\text{R} \propto V_\text{tot} a_\text{R},\quad \langle R\rangle \sim V_\text{tot} \langle r\rangle, \quad N_\text{tot}\sim V_\text{tot} n_0$, which gives us an extensive Fisher information:
\begin{equation*}
\mathcal{I}_{\lambda}^\text{ext} = \frac{V_\text{tot}\left(A_{P}\vec{u}_R\cdot \vec{u}_p\right)^2}{\sigma_{0}^2\langle r \rangle/a_{R} + n_0^{-1}\sigma^2_\text{shot} + V_\text{tot}\sigma^2_\text{enc} }
\end{equation*}

This expression has several informative limits. When particles are abundant, $n_0$ large, decoder noise and (mis)alignment dominate the information. Here, if decoder gain $a_R$ is also large, then alignment is the key factor tuning the amount of information transmission:
\begin{align*}
\mathcal{I}_\lambda^{a_R,n_0\rightarrow\infty} &= 
\frac{A_\text{P}^2\left(\vec{u}_P\cdot\vec{u}_R\right)^2}{\vec{u}_R\cdot\Sigma_\text{enc}\vec{u}_R}\\
&=\frac{A_\text{P}^2\cos^2\theta_{\lambda,R}}{ \left(A_P^2a_\parallel^2\cos^2\theta_{\lambda,R}+a_\perp^2\sin^2\theta_{\lambda,R}+a_\times\sin2\theta_{\lambda,R}\right)^2}\\
\end{align*}

In the second line we have written $\vec{u}_P\cdot\vec{u}_R = \cos\theta_{\lambda,R}$ in terms of the encoder-decoder angle $\theta_{\lambda,R}$. If the encoder noise is mostly along the signal direction and orthogonal directions, $a_\parallel,a_\perp \gg a_\times$, then the information is optimized when $\theta_{\lambda,R} = 0$; if noise moves components in ways that mix directions of the signal, $a_\times$ is large, then $\mathcal{I}_\lambda$ is maximized by $\theta_{\lambda,R}= \arctan(a_\perp^2,-a_\times^2)$: a continuous shift from the aligned case. In both cases the optimal information reads:
\begin{equation*}
\mathcal{I}_\lambda^{\star} = \frac{A_P^2}{A_P^2a_\parallel^2 - a_\times^4/a_\perp^2}
\end{equation*}
Which agrees with the intuition that the information on $\lambda$ is bounded by the intrinsic variance given the encoding process. 

If particles are not abundant, then shot noise dominates. Here, encoder - decoder alignment is still beneficial, though the alignment of the decoder with shot-noise can change the optimal alignment,

\begin{equation*}
\mathcal{I}_\lambda^\text{shot} = A_\text{P}^2V_\text{tot}n_0\frac{\left(\vec{u_R}\cdot\vec{u}_P\right)^2}{\operatorname{Var}(\vec{u}_R)}\\    
\end{equation*}. 

Here the optimal choice of the $\vec{u}_R$ is  $\Sigma_P^{-1}\vec{u}_P$, which in general will not give the encoder-decoder alignment we found above or a slight variation on it. This is because $\Sigma_P$ weighs components differently based on their abundances, distorting the signal away from rare species, which in other cases may be heavily weighted. In this case the optimal information approaches the counting bound:
\begin{equation*}
\mathcal{I}_\lambda^{\text{shot},\star} = A_\text{P}^2V_\text{tot}n_0^{-1}\vec{u}_P\Sigma^{-1}_P\vec{u}_P\\    
\end{equation*}. 

In summary, we have shown that the Fisher information on an environmental signal $\lambda$ is maximized by encoder-decoder alignment. In large, thermodynamic systems this optimization has a simple interpretation of `align the decoder to the signal'. In small systems, or systems with encoder noise, this optimum is different. When encoder noise dominates, the optimal decoder is a continuous shift from simple alignment picture. When shot noise dominates, the optimal decoder is completely different. Below we compute the Fisher - Information for a few prototypical decoders.\\



\noindent\textit{Fisher Information in minimal decoders} -- What makes an encoder $P(\mathbf{n};\lambda)$ geometrically aligned to a decoder $R[P(\mathbf{n})]$? As a motivating example we consider exponential encoder $P(\mathbf{n};\lambda) = \lambda e^{-\lambda n}$ on a continuous trait space. The encoder's variation with the environment as:
\begin{equation*}
\frac{\partial P(n)}{\partial\lambda} = \left(1 -n\lambda\right)e^{-\lambda n}
\end{equation*}
here we notice that $\partial_\lambda P(n) >0$ for $n < 1/\lambda$ and $\partial_\lambda P(n) < 0$ for $n > 1/\lambda$ -- some values of $n$ rise with the $\lambda$, while others fall. A good decoder should consistently map components that rise with $\lambda$ to one sign of output, and those that fall with $\lambda$ to the opposite sign. A simple non-trivial decoder is a linear decoder of the form $R_\text{linear}[P(\mathbf{n})] = \sum_n c n P(n)$. The response in a continuum limit is
\begin{align*}
\int\frac{\delta R}{\delta P_\lambda(\mathbf{n})} \cdot \frac{\partial P_{\lambda}(\mathbf{n})}{\partial\lambda} dn = -\frac{c}{\lambda^2}
\end{align*}
the response scales linearly with the `amplification' $c \equiv V_\text{tot}a_0$, and large $\lambda$ (steep exponentials) have a small response. 

The decoder noise is the variance in response $R$ for a fixed $P(\mathbf{n})$, and can be computed once a physical model is known;  as above we assume that $\sigma^2_\text{dec} = \langle r\rangle V_\text{tot}\sigma_0^2$, which gives the decoder noise as
\begin{align*}
\sigma^2_\text{dec,linear}
&=\sigma_0^2\frac{c}{\lambda}
\end{align*}

Encoder noise is the variance in the mean response given fluctuations in $P(\mathbf{n})$. Shot noise on an exponential distribution is just proportional to the raw counts, and the noise takes the form
\begin{align*}
\sigma_\text{enc}^2 &= \frac{c^2}{N_\text{total}}\int dn \left(n^2 \lambda e^{-\lambda n}\right)\\
&=\frac{c^2}{\lambda^2 n_0V_\text{tot}}
\end{align*}

Putting this all together, the information in a linear decoder is
\begin{align*}
\mathcal{I}_{\lambda}^\text{linear} &= \frac{c n_0 V_\text{tot}}{\lambda^2\left(c + V_\text{tot}n_0\lambda\sigma_0^2\right)}\\
&=\frac{a_0n_0V_\text{tot}}{\lambda^2\left(a_0+n_0\lambda\sigma_0^2\right)}
\end{align*}

\noindent\textit{Exponential decoders} -- Optimality in a noiseless setting requires matching the decoders response $\frac{\delta R}{\delta P(\mathbf{n})}$ to the encoders response $\partial_\lambda P(\mathbf{n})$. This translates into a simple heuristic of matching the signs of these two terms consistently everywhere: consistently amplify traits that move in one direction with the environment, and suppress those that move in the opposite direction. 

While this logic suggests that an optimal decoder is a step function that aligns precisely at $n^*: \partial_\lambda P(\mathbf{n}) = 0$, this decoder can only be optimal at a point. An exponential decoder of the form $R_\text{exponential} = V_\text{tot}a_0\int dn e^{\alpha n}P(n)$ has the total response:
\begin{align*}
\left(\frac{\delta R_\text{exponential}}{\delta P(\mathbf{n})} \cdot \frac{\partial P(\mathbf{n})}{\partial\lambda}\right) &= V_\text{tot}a_0\int{dn  e^{n\left(\alpha-\lambda\right)}\left(1-n \lambda\right)}\\
\\&=-\frac{V_\text{tot}a_0\alpha}{(\alpha-\lambda)^2}
\end{align*}
the expression here diverges as $\alpha \rightarrow\lambda$: a quickly decaying exponential needs more amplification near the tails, and this amplification is formally unbounded. The readout noise is again the mean response:
\begin{equation*}
\sigma_\text{dec,expon}^2 = \frac{V_\text{tot}a_0\sigma_0^2\lambda}{\lambda-\alpha}
\end{equation*}
The encoder noise is the same covariance of inputs, as amplified by the exponential decoder,
\begin{align*}
\sigma^{2}_\text{enc} &= \frac{V_\text{tot}a_0^2}{n_0}\int  e^{n(2\alpha-\lambda)}  dn\\
&= \frac{a_0^2\lambda V_\text{tot}}{n_0(\lambda-2\alpha)}
\end{align*}
Putting these together we get the information in an exponential decoder
\begin{equation*}
\mathcal{I}_\lambda^\text{expon} = \frac{n_0a_0V_\text{tot}\alpha^2\lambda\left(2\alpha-\lambda\right)}{(\alpha-\lambda)^4(a_0\lambda^2 + n_0(\lambda-2\alpha)\sigma_0^2)}
\end{equation*}
Examining the information of a linear and exponential decoder side by side, we see that, when particles are abundant, $n_0\to\infty$, the exponential decoder outperforms the linear decoder when amplification is large, $\alpha \approx \lambda$. When shot noise dominates, $n_0 \to 1$, the linear decoder outperforms the exponential.
   
\noindent\textbf{Non-linear decoders} --- Finally, we investigate non-linear decoders, e.g $R[P(\mathbf{n})] = \sum_n f(n)P(n) $ for some nonlinear function $f(n)$. To explore what properties of $f(n)$ make good decoders, we expand the total response about the mean of $P(\mathbf{n})$
\begin{align*}
R[P(\mathbf{n};\lambda)] &= \langle f(\mathbf{n})\rangle_{P(\mathbf{n})}\\
\\&= \left\langle f(\mu) + \left(\frac{\partial f(n)}{\partial\mu}\right)\left(n-\mu\right)+ \frac{1}{2}\left(\frac{\partial^2 f}{\partial\mu^2}\right)\left(n-\mu\right)^2 +\mathcal{O}((n-\mu)^3)\right\rangle_{P(\mathbf{n})}\\
&= f(\mu) + \frac{1}{2}\left(\frac{\partial^2 f}{\partial\mu^2}\right)\sigma^2 + \dots
\end{align*}
where $\mu = \sum_n n P(n)$, $\sigma^2 = \sum_n (n-\mu)^2P(n)$. The second term depends on both the variance and the convexity of $f(\mathbf{n})$. If $f(\mathbf{n})$ is convex, contours of constant response trace curves where variance decreases with mean. If $f(\mathbf{n})$ is concave, these contours trace curves where variance increases with the mean. 

The mean and variance of exponential distributions increase together. In the plane of $\mu-\sigma^2$, exponential distributions fall on a line $\sigma^2(\mu) \propto \mu^2$. Our above response can be re-written in terms of distribution moments.
\begin{equation*}
\left(\frac{\delta R}{\delta P(\mathbf{n})} \cdot \frac{\partial P(\mathbf{n})}{\partial\lambda}\right) = -\frac{1}{\lambda^2}\left(\frac{\partial f(\mu)}{\partial\mu} + \frac{1}{\lambda}\frac{\partial^2 f(\mu)}{\partial\mu^2} +\frac{3}{2\lambda^2}\frac{\partial^3 f(\mu)}{\partial\mu^3}\right) +\cdots
\end{equation*}
which shows that for exponential distributions, convexity ($\partial_\mu^2 f >0$) is desirable and increases the scale of the response. If variance decreased with the mean, a concave decoder would instead be desirable. For this task, a linear decoder is better than a concave decoder, but worse than a convex non-linear decoder.

\noindent\textit{Noise of a non - linear decoder}: The decoder noise can be computed as before, assuming $\sigma_\text{dec}^2 \propto \langle R\rangle \sigma_0^2$. Using the above expansion we obtain the noise:
\begin{align*}
\sigma_\text{dec}^2 &= \langle f(\mathbf{n})\rangle\sigma_0^2\\
&= \left(f(\mu) + \frac{1}{2}\left(\frac{\partial^2f}{\partial\mu^2}\right)\sigma^2+\dots\right)\sigma_0^2
\end{align*}

The shot noise can be computed as before, pushing the response through the covariance matrix; for an exponentially distributed inputs this 
\begin{align*}
\sigma_\text{shot}^2 &= \frac{1}{N_\text{tot}} \bigl\langle f(n)^2\bigr\rangle - \left\langle f(n)\right\rangle^2  \\
&= \frac{1}{\lambda^2}\left(\frac{\partial f}{\partial\mu} +\frac{1}{\lambda}\frac{\partial^2f}{\partial\mu^2}\right)^2+\frac{1}{\lambda^4}\left(\frac{\partial^2f}{\partial\mu^2}\right)^2 + \dots
\end{align*}
Defining $m\equiv\partial_\mu f,\quad \kappa \equiv \partial_{\mu}^2f$, we combine all of these terms into the information:
\begin{equation*}
\mathcal{I}_\lambda^\text{nonlinear} =  \frac{N_\text{tot}(\kappa + m\lambda)^2}{\lambda^2\left((\kappa+m\lambda)^2 + N_\text{tot}\sigma_0^2\langle
  f(\mathbf{n})\rangle\lambda^4\right)}
\end{equation*}
Here, as $N_\text{tot}\to\infty$ increasing $\kappa$ will only increase the information until saturation. In the opposites limit, $\kappa$ provides no benefit since it cancels with the factors in the shot noise.

\section{Simple Binding}\label{sec:binding}  \noindent\textbf{Model} --- We model a system consisting of singly valent $B$ binders and multiply valent input $A_n$ with a conventional mass-action approach. The dynamical equations dictating chemical equilibrium are:
\begin{align}
\frac{dA_{n,i}}{dt} &= B_\text{free}k_\text{on}A_{n,i-1} + k_\text{off}A_{n,i+1} -A_{n,i}\left(k_\text{off} + B_\text{free}k_\text{on}\right)\\
\frac{dB_\text{free}}{dt}&= k_\text{off}\left(B_\text{tot}-B_\text{free}\right)\left(\sum_{n=1}^{N_\text{sites}}\sum_{i=1}^{N_\text{sites}}A_{n,i}\right) - k_\text{on}B_\text{free}\left(A_\text{tot} - \sum_{n=1}^{N_\text{sites}}\sum_{i=1}^{N_\text{sites}}A_{n,i}\right)
\end{align}
where $n$ denotes the valency of input molecules and $i$ denotes the binder occupancy on that molecule, e.g $A_{4,2}$ is the 4-valent input bound to 2 binder molecules. 

The steady-state solution is obtained by noting that the dynamics of $B$ only depend on $A_\text{bound} \equiv \sum_{n=1}^{N_\text{sites}}\sum_{i=1}^{N_\text{sites}}A_{n,i}=A_\text{tot}\mu_1$ and the free binder $B_\text{free}$. Independent binding statistics give  $B_\text{tot} - B_\text{free} = B_\text{tot}p_\text{bind}\sum_{n=1}^{N_\text{sites}}\sum_{i}^{N_\text{sites}} A_{n,i} = B_\text{tot}p_\text{bind}A_\text{tot}\mu_1$, where $p_\text{bind}$ is the per-site binding probability that is to be determined. Thus $B_\text{free}$ can be written as
\begin{align*}
B_\text{free} = B_\text{tot}\left(1- p_{\text{bind}}A_\text{tot}\mu_1\right)
\end{align*}
We can obtain the binding probability $p_\text{bind}$ in terms of an equilibrium constant $k_\text{D} = \frac{k_\text{off}}{k_\text{on}}$
\begin{equation*}
p_\text{bind} = \frac{B_\text{free}/k_{D}}{1+B_\text{free}/k_{D}}
\end{equation*}

Combining these two equations we can solve for the steady-state $B_\text{free}$ and the resulting occupancies of input molecules. $B_\text{free}$ is given by the roots of a quadratic equation:
\begin{equation}\label{eqn:BFREE}
B_\text{free}^{\star} = B_\text{tot}\left(1-A_\text{tot}-\mu_1\right)- k_{D} \pm \sqrt{( 4B_\text{tot}k_{D} + \left(k_D + B_\text{tot}\left(A_\text{tot}\mu_1 - 1\right)\right)^2}
\end{equation}

At any given point in parameter space, one of these roots is physical, satisfying $0 \le B_\text{free} \le B_\text{tot}$. The free binder fraction $B_\text{free} / B_\text{tot}$ is only influenced by the mean site availability $\mu_1$ and total input concentration $A_\text{tot}$. The input-binder complex concentrations can be calculated from independent binding statistics: 
\begin{align*}
A_{n,i}^{\star} &= A_{n,\text{tot}}{n\choose i}(p_\text{bind})^i (1-p_\text{bind})^{n-i}\\
p_{n,i}^{\star} &={n\choose i}(p_\text{bind})^i (1-p_\text{bind})^{n-i}
\end{align*}
Here $p_{n,i}^{\star}$ is the steady-state probability of the input with $n$ sites being bound by $i$ binders, and $A_{n,tot}$ is the input concentration of input with $n$ total sites. From these expressions we obtain the complex occupancy $C$, 
\begin{align*}
CA_\text{tot} &= A_\text{tot} - \sum_{i=0}^{N_\text{sites}}A_{0,i}\\
&= A_\text{tot} - A_{0,\text{tot}} - \sum_{n=1}^{N_\text{sites}}A_{n,\text{tot}}\left(1-\sum_{i=1}^{N_\text{sites}}p_{n,i}^\star\right)\\
&= A_\text{tot} - \sum_{n=1}^{N_\text{sites}}A_{n,\text{tot}}\left(1-p_\text{bind}\right)^i\\
R_\text{binding} &= 1 -\sum_{n=1}^{N_\text{sites}} P(n)\left(1-p_\text{bind}\right)^n
\end{align*}
Here the $P(n)$ is the normalized input distribution. From the above expression we can read off the curvature of the single-molecule response, $\partial^2_{n}R_\text{binding} = -(1-p_\text{bind})^n\log(1-p_\text{bind})^2 <0$; $R_\text{binding}$ is always concave.


\section{Phase-Separation}\label{sec:phase}
\subsection*{Analysis of Mean Field Theory}
In the main text we present phase diagrams characterized by regions of thermodynamic instability, multistability, and critical points. Below we derive instability and critical point conditions for a multi-component fluid, and conditions for multistability that permit efficient calculations of phase-boundaries and volumes. 

Our system consists of an input molecules $A_n$ with $0 - N_\text{sites}$ chemically equivalent sites through which it interacts with a Binder molecule $B$, which itself has valency $n_{B}$. Each site on the binder interacts with the energy $J_{B}/n_{B}$. We assume that the energy density of interactions scales with the number of available sites, and write the mean-field free energy density below.
\begin{align}\label{eqn:fhenergy}
f_\text{sys}=\frac{F_\text{sys}}{k_BT} = & \sum_{n=0}^{N_\text{sites}} \left(\phi_{A_n}\log{\phi_{A_n}} + \frac{n J_{B}}{k_B T} \phi_{A_n}\phi_B\right)\nonumber\\ &+\phi_\text{B} \log \phi_\text{B} + \left(1-\sum_{n=0}^{N_\text{sites}}\phi_{A_n} - \phi_B\right) \log \left(1-\sum_{n=0}^{N_\text{sites}}\phi_{A_n} - \phi_B\right)    
\end{align}
Here $\phi_B$ is the volume fraction of binder molecules, $\phi_{A_i}$ is the volume fraction of the $i^{th}$ input molecules.  

\noindent\textit{Thermodynamic Instability}: --- A homogeneous composition $\lbrace\vec\phi\rbrace=\lbrace \phi_{B},\phi_{A_0}\dots\phi_{A_{N}}\rbrace$ is unstable when the free energy density can locally decrease. This condition implies that there is a direction in composition space along which the derivative of the free energy is negative, or equivalently that the Hessian matrix $H_{i,j} = \partial_{\phi_i}\partial_{\phi_j} f_\text{sys}$ has at least one negative eigenvalue. For a Flory-Huggins free energy density the Hessian is:
\begin{equation*}
H_{i,j} = H_{j,i} = \underbrace{\frac{1}{1-\phi_B-\phi_{A,\text{tot}}}+\frac{\delta_{i,j}}{\phi_i}}_{K_{i,j}} + J_{i,j}
\end{equation*}
The first two terms are always positive. For our system the interaction term takes the form $J_{i,j} = J_{j,i} = i J_{B}/k_BT$ for $n \leq N_s$,\quad $j=N_s+1$ (or vice versa). Together these terms define a positive semi-definite matrix $K_{i,j}$ with a perturbation $J_{i,j}$. The Hessian is unstable when the lowest eigenvalue first becomes zero, or when the inverse of the Hessian first becomes undefined.  

This condition of marginal stability can be computed via the the Woodbury matrix identity. We consider our Hessian as a low-rank perturbation $\mathbf{J}$ to a matrix $\mathbf{K}$ with a well - defined inverse:
\begin{equation}
\mathbf{H}^{-1} = \mathbf{K}^{-1} + \mathbf{K}^{-1}\mathbf{U}\left(\mathbf{C}^{-1} + \mathbf{U}^{T}\mathbf{K}^{-1}\mathbf{V}\right)^{-1}\mathbf{V}\mathbf{K}^{-1}
\end{equation}  
where matrices $C$,$U$,$V$ are a suitable decomposition of our interaction matrix defined via $\mathbf{J} = \mathbf{UCV}$. To decompose our interactions we note that the interactions only have entries in the binder column, and that these increase with input valency. The relevant matrices to get this structure is a $2\times \left(N_\text{sites}+2\right)$ matrix: 
\begin{equation*}
U = V= \begin{pmatrix}
0 & 0 & 0 & 0 & 0 &0 & 0 &1\\
0 & 1 & 2 & 3& 4 &5& 6 & 0
\end{pmatrix}\end{equation*}
and $C = \frac{J_\text{B}}{k_{B}T}\begin{pmatrix}0 & 1\\1&0 \end{pmatrix}$.

The inverse of $\mathbf{K}$ is $ K_{i,j}^{-1}=\delta_{i,j}\phi_i-\phi_i\phi_j$, which is always defined. From the Woodbury identity it can be seen that $\mathbf{H}^{-1}$ only exists if $\left(\mathbf{C}^{-1} + \mathbf{U}^T\mathbf{K}^{-1}\mathbf{V}\right)^{-1}$ exists. Together we can construct $M=U^{T}K^{-1}V$, a $2\times 2$ matrix that will determine the stability of the Hessian
\begin{align*}
M =U^TK^{-1}V &= \begin{pmatrix}
\phi_{B}(1-\phi_B) & -\phi_B\mu_1\phi_\text{A,tot}\\-\phi_B\mu_1\phi_\text{A,tot}&\phi_\text{A,tot}\left(\mu_2 - \phi_{A,\text{tot}}\mu_1^2\right)
\end{pmatrix}\\
C^{-1} &= \frac{k_{B}T}{J_\text{B}}\begin{pmatrix}
0&1\\1&0    
\end{pmatrix}
\end{align*}
where $\mu_1 = \frac{1}{\phi_{A,\text{tot}}}\sum_{n}^{N_\text{sites}}n\phi_{A_n}$, $\mu_2 =\frac{1}{\phi_{A,\text{tot}}} \sum_{n}^{N_\text{sites}}n^2\phi_{A_n}$ are the mean and second moment of the input distribution, respectively. The homogeneous mixture is at the limit of instability when the determinant of $C^{-1} - M$ is zero which gives the spinodal condition:
\begin{equation}\label{eqn:spincond}
\phi_B\left(1-\phi_B\right)\phi_{A,\text{tot}}\left(\mu_2-\phi_{A,\text{tot}}\mu_1^2\right)= \left(\phi_\text{A,tot}\mu_1\phi_B +\frac{k_{B}T}{J_{B}}\right)^2  
\end{equation}
For a given input distribution the physically relevant solutions to ~\ref{eqn:spincond} give the spinodal line in terms of a relationship between $\phi_{A,\text{tot}}$ and $\phi_{B}$. Together this analysis suggests that moments higher than second order cannot be discriminated \textit{on the basis of stability alone}. It is still possible to discriminate distributions via the volumes, phase - compositions, and in the binodal region. \\

\noindent\textbf{Critical Points}: --- Critical points occur when, for a given composition $\vec\phi$, the following conditions are satisfied: 
\begin{align*}
\partial_{\phi_i}f_\text{sys}  &= 0\\
\partial_{\phi_i}\partial_{\phi_j} f_\text{sys}  &= 0 \qquad\forall \quad i,j\\
\partial_{\phi_i}\partial_{\phi_j}\partial_{\phi_k}f_\text{sys}  &= 0  \qquad\forall \quad i,j,k
\end{align*}
where the first condition is that the system is at an extrema of the free energy, the second derivative condition is that of marginal thermodynamic stability we analyzed above. The final equation represents, informally, a `zero-skewness' condition on the composition, where the direction of the derivative is taken to be the direction of marginal stability. Since our free-energy incorporates only coupling up to second order, $\frac{\partial^3f_\text{sys}}{\partial\phi_i\partial\phi_j\partial\phi_k} = \delta_{ij}\delta_{jk}\left(\frac{1}{\left(1-\phi_B-\phi_{A,\text{tot}}\right)^2}-\frac{1}{\phi_i^2}\right)$ and the third derivative condition takes a compact form $\delta f_\text{sys}(\epsilon) = \epsilon^3\sum_i\left(V^0_i\right)^3\frac{\partial^3 f}{\partial\phi_i^3} = 0$, where $V_i^0$ is $i^{th}$ component of the marginally stable eigenvector. By the condition $H\cdot V^0 = 0$, the eigenvector $V^{0}$ has the form:
\begin{align*}
V_n^0 = \frac{J_{B}}{k_{B}T}\phi_{A_n}\left(n - n^{\star}\right)V^{0}_B \quad n\neq B\\
n^{\star} = \frac{\phi_{A,\text{tot}}\mu_1+J_{B}^{-1}}{1-\phi_{B}}
\end{align*}
where $V^0_{B}$ is the direction of the $B$ molecule, which we can set as $V^0_{B}=1$. The critical point condition becomes
\begin{align}
\sum_n^{N_\text{sites}+1}\left(V^0_n\right)^3\left(\frac{1}{\phi_{A,n}^2} - \frac{1}{\left(1-\phi_{A,\text{tot}}-\phi_B\right)^2} \right) &= 0\\
\sum_{n}^{N_\text{sites}}\left(\frac{\left(J_{B}\phi_{A_i}\left(n-n^{\star}\right)\right)^3}{\phi_{A_n}^2}  \right) &= \left(\frac{1}{\phi_B^2} - \sum_{n}^{N_\text{sites}+1}\frac{\left(V_n^0\right)^3}{\left(1-\phi_{A,\text{tot}}-\phi_B\right)^2} \right)\\
J_{B}^3\phi_{A,tot}\left\langle \left(n - n^{\star}\right)^3\right\rangle &=\frac{1}{\phi_B^2} + \frac{\left(1-\phi_{B}-\phi_{A}\right)\left(1+J_{B}\phi_{A,\text{tot}}\mu_1\right)}{(1-\phi_B)^3(1-\phi_{A,tot}-\phi_{B})^2}\\
\end{align}
Where the average is taken over the input distribution $\langle \dots\rangle = \sum_{i}^{N_S}{\phi_i \dots}$. Together with the spinodal condition this can be used locate critical points in the plane of $\phi_{A,\text{tot}}$ and $\phi_{B}$. The critical point condition depends on moments up to order $3$ in the input distribution, while the spinodal condition depend on moments up to order $2$. This suggests that two input distributions with the same mean and variance can have different critical points. Since binodals and spinodals meet at critical points, this shows that these distributions can have different binodal regions even though their spinodals are the same. \\

\noindent\textbf{Locating Binodal Lines}: --- We have shown that the spinodal does not depend on details of the input distribution past the second moment, and critical points do not depend on details higher than the third moment. To efficiently compute phase boundaries, we note that at coexistence, the composition of the majority `parent' phase (which we will write as $\phi^\text{dil}$, without loss of generality) must match the input composition, $\phi^\text{inp} = \phi_{A,\text{tot}}P(A_n)$, where $P(A_n)$ is distribution of inputs. This condition specifies the input distribution along the coexistence curve:
\begin{equation*}
\phi_{A_n}^{\text{dil}} = P(n) \phi_\text{A,tot}^{\text{dil}} 
\end{equation*}
Now we use the thermodynamic relation that any two phases at co-existence have equal chemical potentials to obtain the partition coefficients $K_i = \frac{\phi_{A_n}^{\text{dense}}}{\phi_{A_n}^\text{dilute}}$ of the each input species between the dense and dilute phases
\begin{align*}
\log\left(\frac{\phi_{A_n}^\text{dense}}{\phi_{A_n}^\text{dilute}}\right) &=  n \underbrace{J_{B} \left(\phi_B^{\text{dense}} - \phi_B^\text{dilute}\right)}_{\gamma}+\underbrace{\log\left(\frac{1-\phi_{A,\text{tot}}^\text{dense}-\phi_B^\text{dense}}{1-\phi_\text{A,tot}^\text{dilute}-\phi_B^\text{dilute}}\right)}_{\alpha} \\
\left(\frac{\phi^\text{dense}_i}{\phi_i^{\text{dilute}}}\right) &= e^{n\gamma+\alpha}
\end{align*}
Now we insert the input distribution for the dilute phase, and rearrange to obtain the total dense phase composition as a function of $\gamma = J_\text{B}\left(\phi_{B,\text{dense}} - \phi_{B,\text{inp}}\right)$
\begin{align*}
\phi_{A,\text{tot}}^\text{dense} &= 
\sum_n\phi_{A_n}^{\text{dense}}\\
&=\sum_n\phi_{A_n}^\text{dilute}e^{n\gamma+\alpha}\\
&= \phi_{A,\text{tot}}^{\text{dilute}}e^{\alpha}\underbrace{\sum_{n}e^{n\gamma}P(n)}_{Z(\gamma)}
\end{align*}
The distribution of input molecules across the dense phase, $Q(n)$, is obtained by suitably normalizing the individual compositions:
\begin{align*}
Q(n) &= \frac{\phi_{A_n}^{\text{dense}}}{\phi_{A,\text{tot}}^\text{dense}}\\
&= \frac{e^{n\gamma+\alpha} P(n)}{Z(\gamma)}
\end{align*}
From this relation it can be seen that, at the phase boundary, the dense-phase input distribution is an \textit{exponentially tilted} version of the parent input distribution, with strength depending on the degree of binder and total input enrichment. Finally, the partition ratio of inputs $e^{\alpha}$ can be eliminated via:
\begin{align*}
e^{\alpha} &= \frac{\phi_{A,\text{tot}}^\text{dense}}{\phi_{A,\text{tot}}^\text{dil}}\\
&= \frac{1-\phi_{B}^\text{dense}}{\phi_{A,\text{tot}}^\text{dil}(Z(\gamma)-1)+1-\phi_{B}^\text{dil}}
\end{align*}

All together these relations suggests an efficient method for locating the coexistence curve:
\begin{itemize}
    \item Specify a input distribution $P(n)$
    \item Choose a candidate binodal point $(\phi_{B}^\text{dilute},\phi_{A,\text{total}}^\text{dilute})$
    \item Compute $F(\vec{\phi^\text{dilute}})$, $\mu_B\left(\vec{\phi^\text{dilute}}\right)$
    \item Iterate through $\gamma \in [-J_B,J_B]$ values, defining $Q(n)=P(n)e^{i\gamma+\alpha}/Z(\gamma)$, $\phi_{B}^\text{dense} = \phi_{B}^\text{dilute} + \frac{\gamma}{J_{B}}$
    \item Compute $F(\phi^\text{dense}) = F(\phi^\text{dilute})$ and $\mu_B(\vec{\phi^\text{dense}}) = \mu_B(\vec{\phi^\text{dilute}})$, and check if there is a $\gamma$ that satisfies both of these equalities. If so, the candidate point is on the binodal. 
\end{itemize}

This reduces the problem of optimizing $2\times \left(N_\text{sites}+2\right)$ coexisting compositions to scanning for values of $\gamma$ and computing chemical potentials and free energies. 

This relation between the input distribution in the dilute and dense phases at the phase-boundary has the form of a moment-generating function $Z(\gamma)$ for the input distribution. The evaluation at $\gamma = 0$ returns the moments of the parent distribution. To compare two distributions with different input moments, we compute the total site fraction $\mu_{1,Q} = \sum_{i}^{N_\text{sites}}nQ(n)$:
\begin{align*}
\mu_{1,Q}  &= \frac{\partial Z}{\partial\gamma}\\
&=\mu_{1,P} + \mu_{2,P} \gamma + \mu_{3,P}\gamma^2 +\mu_{4,P} \gamma^3 +\mathcal{O}\left(\gamma^4\right)
\end{align*}
where the $\mu_{k,P}$ are moments of the parent distribution. By the structure of these equations the $k^{th}$ moment of the parent distribution will be related to $j-k$ moments in the $Q$ distribution. This shows that near the critical point (small $\gamma$), distributions are only distinguishable by low order moments. Away from the critical point all $\gamma$ contribute to the shape of the binodal, albeit at higher and higher orders. \\

\noindent\textbf{Phase Coexistence beyond the phase boundary}: --- When the system is not at a critical point, thermodynamically unstable, or exactly at a phase boundary we directly minimize $f_\text{sys}$ subject to mass constraints and chemical potential equality. Thus the free energy $F_\text{sys} = V_\text{total}f_\text{sys} = V_{I}f(\vec{\phi_1}) + V_{II}f(\vec{\phi_{II}})$ subject to the mass and volume constraints:
\begin{align*}
\vec{\phi_{I}} V_I + \vec{\phi_{II}}V_{II} &= V_\text{total}\vec{\phi_{\text{inp}}}\\
V_I + V_{II} &= V_\text{total}\\
\end{align*}
where $V_i$ and $\phi_i$ are the volume and composition of the $i^{th}$ phase. Minimization of this free energy implies equality of chemical potentials $\mu_i = \left(\frac{\partial f}{\partial\phi}\right)_{\phi_i}$ for all phases. This implies a relation analogous to the exponential tilting we leveraged to compute the binodal:
\begin{align*}
\phi_{A_n}^\text{dense}  &= e^{n\gamma+\alpha}\phi_{A_n}^\text{dil}\\
Q(n)\phi_{A,\text{tot}}^\text{dense} &= P(n)\phi_{A_n,\text{tot}}^\text{dil}e^{n\gamma+\alpha}
\end{align*}
where in contrast to the previous example the composition of both phases are undetermined. With mass constraints the dilute phase composition can be eliminated, $\phi_{A_n}^\text{dil} = \frac{\phi_{A_n}^\text{inp}+v_\text{dense}\phi_{A_n}^\text{dense}}{1-v_\text{dense}}$. This resolves into minimization of $f_\text{sys}$ over variables $\gamma$,$\Delta\phi_{A}^\text{tot}$, and $v_\text{dense}$. 


\subsection*{Fluctuations and Information}
\noindent\textbf{Calculating correlation lengths}: --- To define the response in the dilute phase, we calculate the size of correlated fluctuations in composition. Below we derive the correlation length as eigenvalues of the Hessian evaluated at the dilute phase minima. 

We begin by adding gradient terms to the free energy density in Equation ~\ref{eqn:fhenergy}, the inhomogeneous free energy takes the form:
\begin{align*}
\frac{F_\text{sys}}{k_BT} =  \int d^dx\left( f_\text{sys}(\vec{\phi}) + \sum_{i,j}\kappa_{i,j}\nabla\phi_i\nabla\phi_j\right)
\end{align*}
where $\kappa_{i,j} = \kappa_{j,i}$ are terms that penalize spatial gradients and parameterize the length  scale of interactions. For simplicity we assume that $\kappa_{i,j} \propto J_{i,j}$, since all molecules are of the same size. To calculate the correlation length we consider the response  $\delta\vec{\phi}$ a small compositional perturbation, and linearize the system about its energy minimum. Writing the response in matrix form:
\begin{equation*}
\delta\vec{\phi} = \left(k^2 - \mathbf{K}^{-1}\mathbf{H}\right)^{-1}
\end{equation*}
Where $\mathbf{K}$ is the matrix of interfacial coefficients, $\mathbf{H}$ is the Hessian we defined earlier, and $k$ is the wavenumber. We identify the correlation lengths with the inverse eigenvalues of scaled Hessian $K^{-1}H$, e.g $\xi_i = \alpha_i^{-1/2}$ where $\alpha_i$ is the $i^\text{th}$ eigenvalue. We focus on the largest correlation length as the relevant response, $\xi \equiv \alpha_\text{min}^{-1/2}$. 

\noindent\textit{Sensitivity in the dilute phase}: The correlation volume $\xi^d$ in the dilute phase changes as a function of the input distribution. This sensitivity is:
\begin{align*}
\chi_{R,\lambda}^\text{dilute} &= \frac{\partial \xi^d}{\partial\lambda}\\
\end{align*}
where $\lambda$ parametrizes the change in the distribution, like the rate-parameter for exponential distributions. Defining the change in the smallest eigenvalue of the Hessian $\chi_\text{min}$ with $\lambda$ via $\kappa = -\frac{1}{2}\frac{\partial\alpha_\text{min}}{\partial\lambda}$ allows us to rewrite the $\chi_{R,\lambda}^\text{dil}$ as
\begin{align*}
\chi_{R,\lambda}^\text{dil} &= \frac{\partial}{\partial\lambda}\frac{1}{\alpha_\text{min}^{d/2}}\\
&= d\kappa\alpha_\text{min}^{-(d/2+1)}\\
&=  d\kappa\chi_\text{dil}^{(d/2+1)}
\end{align*}
Here $\chi_\text{dil} = \alpha_\text{min}^{-1}$ sets the scale of the readout, and $\kappa$ measures the rate at which distributional shifts changes the scale of the readout.

\textit{Correlation volume fluctuations} --- The readout noise in the dilute phase, $\chi_{R,R}$, are the typical fluctuations in the correlation volume. The product of two fluctuating quantities can be evaluated in a Gaussian approximation as $\langle\phi(r_1)^2\phi(r_2)^2\rangle = 2\langle\phi(r_1)\phi(r_2)\rangle^2 + \langle\phi^2\rangle^2$; the variance is proportional to the square of the mean. This relation implies that the correlation length fluctuations are suitable powers of the correlation length itself. To obtain the readout noise, we divide by the number of independent correlation volumes
\begin{align*}
\chi_{R,R}^\text{dil} &= \frac{\xi^{d}}{V_\text{tot}/\xi^d}\\
&=\frac{\xi^{2d}}{V_\text{tot}}
\end{align*}

\textit{Droplet volume response}: when two phases co-exist, the response is defined by the volume weighted sum of dilute and dense phase compositions. 
\begin{equation*}
\chi_{R,\lambda}^\text{dense} = V_\text{tot}\frac{\partial v_\text{dense}}{\partial\lambda} + (1-v_\text{dense}) \left(\frac{\partial \xi^d}{\partial\lambda}\right)_{\vec{\phi}^\text{dilute}}
\end{equation*}
the first term is the change in the droplet volume with $\lambda$. 

\textit{Droplet volume fluctuations} --- The noise in the readout when two phases co-exist is given by the fluctuations in droplet volume, $\langle V_\text{dense}^2\rangle$. Phase-equilibria sets $\frac{\partial^2 F}{\partial V_{d}^2} = 0$; one can always exchange material between the dense and dilute phases at their equilibrium compositions. Volume fluctuations either originate from correlated composition fluctuations along the dense phase minima, or size fluctuations driven by the surface tension. Here we only account for volume fluctuations from particle exchange as tension related effects are subdominant for large droplet sizes. The dense phase composition fluctuates around its mean value via
\begin{equation*}
\delta\vec{\phi}_\text{dense} = \chi_\text{dense}\vec{V}_0
\end{equation*}
where $\chi_\text{dense} = 1/\alpha_\text{min}^\text{dense},\vec{V}_0$ are the smallest eigenvalue of the Hessian evaluated at the dense phase composition, and its accompanying eigenvector. Volume fluctuations are obtained via the lever rule:
\begin{equation*}
\delta V_\text{dense} = V_\text{tot}v_\text{dense}\delta\vec\phi
\end{equation*}
These fluctuations are small in most regions of parameter space, except near critical points where the susceptibility diverges and phase volume fluctuations are on the scale of the system size. 

\noindent\textit{Fisher Information} --- Putting together the response and fluctuations in  dense and dilute phases, the Fisher information from the response on $\lambda$ is:
\begin{align*}
\mathcal{I}_{\text{phase},\lambda} &= \frac{\left(\partial_\lambda R_\text{phase}\right)^2}{\sigma_{R}^2}\\
&= \frac{\left(V_\text{tot} \partial_{\lambda}v_\text{dense} + \partial_{\lambda}\xi^d\right)^2}{V_\text{tot}v_\text{dense}\chi_\text{dense}+\xi^{2d}/V_\text{tot}}
\end{align*}
There are several informative limits of the Fisher Information. First, in the dilute phase away from critical points the information reduces to:
\begin{align*}
\mathcal{I}_\lambda^\text{dilute} &= \frac{V_\text{tot}}{\xi^{2d}}\left(\partial_{\lambda}\xi^d\right)^2\\
&= \frac{V_\text{tot}}{\chi_\text{dil}^d}\left(d\kappa\chi_\text{dil}^{d/2+1}\right)^2\\
&= V_\text{tot}d^2\chi_\text{dil}^2\kappa^2
\end{align*}
where we have defined $\kappa$, $\chi_\text{dil}$ in the previous section. The Information scales linearly with the system size, and the sensitivity enters as prefactors that are small far from the critical point. 
Deep in the co-existence region we can neglect the dilute phase terms and the information is:
\begin{align*}
\mathcal{I}_\lambda^\text{dense} &= V_\text{tot}\frac{\left( \partial_{\lambda}v_\text{dense}\right)^2}{v_\text{dense}\chi_\text{dense}}\\
&\approx \frac{V_\text{tot}}{v_\text{dense}\chi_\text{dense}}\frac{\kappa}{\Delta\phi}
\end{align*}
where $\Delta\phi = \phi_\text{dense}-\phi_\text{dilute}$ is the miscibility gap. again the information scales linearly with the volume, but the prefactors are larger than in the dilute phase.  Exactly at the phase boundary as $v_\text{dense}\to0$ the information scales super-linearly. 
\begin{align*}
\mathcal{I}_\lambda^\text{binodal} &= \frac{V_\text{tot}^2}{\chi_\text{dil}^{d}}\frac{\kappa}{\Delta\phi}
\end{align*}
which in practice will always be capped by the noise floor of the substrate.

Finally, near a critical point the susceptibility scales with the system size, making the information in the dilute phase:
\begin{align*}
\mathcal{I}_\lambda^\text{crit} \sim V_\text{tot}^{1+(2-\eta)/d}\kappa
\end{align*}
where $\eta$ is the anomalous scaling exponent of the correlation length. Mean - field exponent of $\eta=0,\quad d=4$ gives $V_\text{tot}^{3/2}$ -- information is amplified near the critical point. 

\section{Lattice Monte-Carlo Simulations}\label{sec:mc}
Our Monte-Carlo simulations are an extension of traditional multicomponent lattice gas simulations. The Hamiltonian of our system is
\begin{equation*}
H = \frac{J_\text{B}}{k_B T}\sum_i b_i +\frac{J_{nn}}{k_B T}\sum_{<i,j>}s_i s_j
\end{equation*}
Here $J_B$ is the energy of bonds in the system, $b_i$ is a variable that is one when a bond exists, and zero otherwise. $J_{nn} = -0.2 k_BT$ is a weak nearest-neighbor interaction energy, and the $s_i$'s are spin variables that are one when a molecule is present, and zero when solvent is present. We include these nearest neighbor interactions to give our droplets a 'tension', and to mimic any generalized hydrophobic force between molecules. Implicit in our Hamiltonian are `bonding constraints', which specify how many bond a particular molecule can have. The input molecules have zero to six total bonds, while the binder molecule has four total bonds. We enforce bonding constraints through our move proposals, which we detail below.\\

\noindent\textbf{Simulation Procedure}: Each Monte Carlo step we propose either a swap move, which exchanges two particles at distinct, non-neighboring sites, or a bond-rearrangement moves that destroys and creates bonds at a particular lattice site.

\noindent\textit{Bond moves}: For bond moves, we choose a random lattice site and a random direction. If a bond exists, we propose a deletion with the Boltzmann probability $P_\text{bond} = e^{\beta J_\text{bond}}$. If there is no bond at that site, and a neighboring molecule is free to bond we propose a bond addition with immediate acceptance. The probability of proposing forward and reverse moves is equal, $P = \frac{2}{z V}$, where $z$ is the lattice coordination number and $V=L^3$ is the volume of the lattice. The factor of two comes from counting the ways to add a specific bond, which for binary interactions is two. The acceptance probability is Metropolis, thus these moves satisfy detailed balance.

\noindent\textit{Swap moves}: We propose swap moves to sample the positional degrees of freedom. We choose two random lattice sites and propose an exchange the particles at these sites. Upon exchange, we randomly reassign bond configurations at each site, subject to availability constraints on neighboring molecules. We propose a given swap with probability $\frac{1}{\Omega_1\Omega_2}$, where $\Omega_i$ is single-site partition function for site $i$: the number of distinct bonding configurations available to that site, a combinatorial function of the available bonds and the valency at site $i$. For example, an input molecule of valency $4$ next to $3$ binders with available sites is assigned $\Omega = 9$ If the move is proposed we randomly assign bonds in the new configuration. 

This scheme satisfies detailed balance because the probability of choosing a particular configuration at a given site at random is the inverse of the single site partition function, and the joint probability of both events is the product of these two weights. We accept the move with a Metropolis probability given by total change in energy. 

We initialize the system in a $L\times L \times L$ lattice with $L = 24$, randomly assigning molecules to sites in accord with composition constraints, and with no initial bonds. We simulate for $1\times10^9$ iterations.\\

\begin{figure}\label{fig:FigS2}
    \centering
    \includegraphics[width=\linewidth]{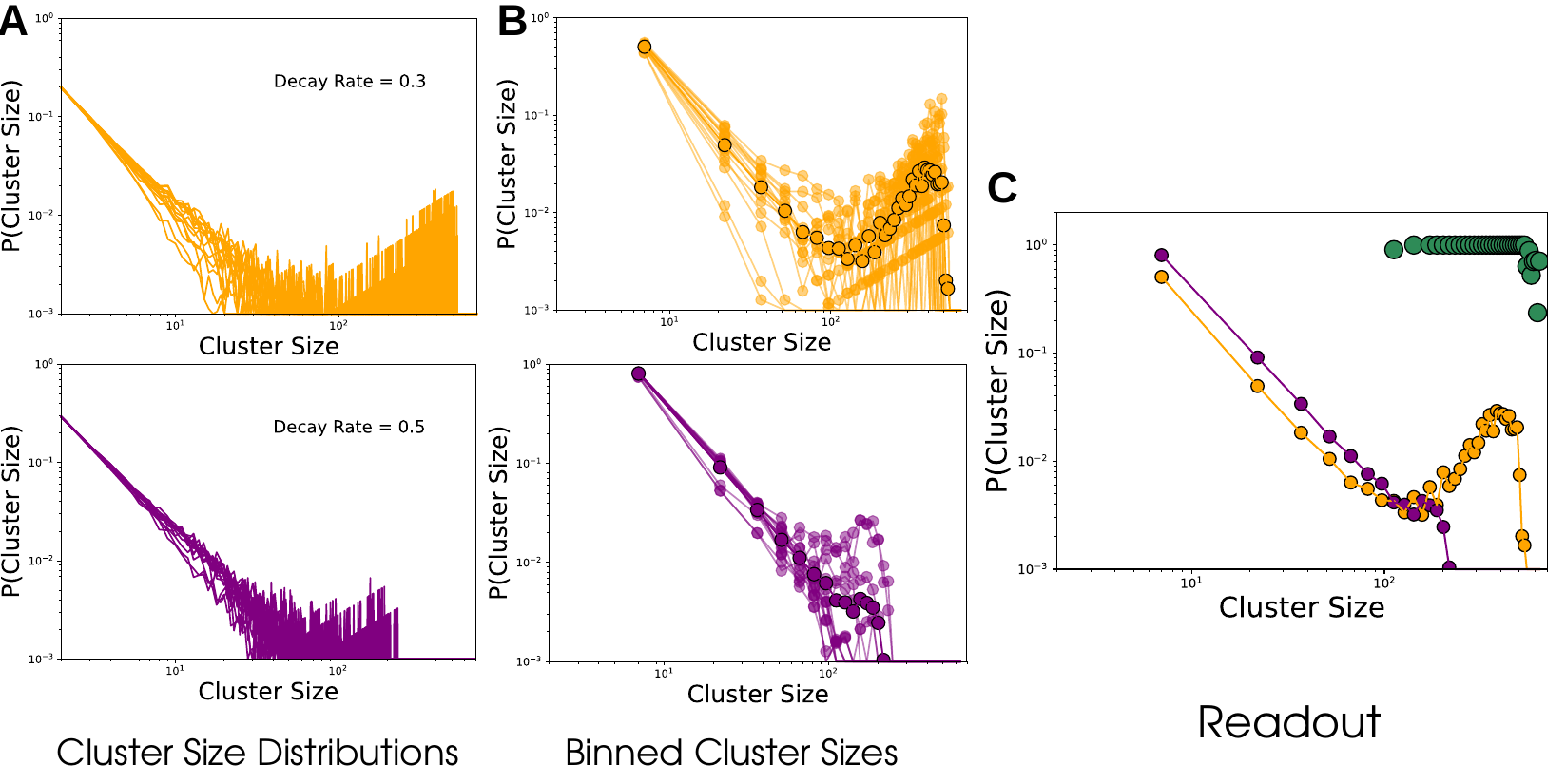}
    \caption{\textbf{Quantifying Monte-Carlo Simulations} --- Workflow for analyzing Monte-Carlo Simulations. (A) Cluster size distributions are extracted from independent simulations. We weigh these distributions by the size of the cluster. Simulation that phase-separate (upper, orange) and remain well-mixed (lower, purple) (B) Cluster sized binned with even spacing of 15. Independent simulations are transparent, opaque is the mean. (C) Mean distributions and optimized readout weights: we optimize the a linear function to separate the mean cluster size distributions. These coefficients are marked as green dots, and are either one for regions where there is a difference between the two distributions, or approximately zero}
\end{figure}

\noindent\textbf{Quantifying Monte-Carlo Simulations}  --- In the main text we present optimized `readout' functions from a series of Monte Carlo Simulations. Our readout function is a weighted linear sum of binned cluster sizes. Below we explain how we obtain the cluster size distribution, optimization of the readout function, and how we located approximate phase boundaries. We illustrate this procedure in Figure ~\ref{fig:FigS2}\\

\textit{Cluster Size Distributions} --- We obtain cluster size distributions from a set of lattice configurations. For a particular configuration we calculate size of  nearest-neighbor cluster of non-solvent molecules (Input and Binder) molecules. We average this distribution over many individual configurations from independent simulations. The resulting pooled cluster size distribution is then binned into logarithmically spaced intervals of $[1,5,10,50,100,500,1000]$. This distribution over $7$ states is then fed to our readout function, which we optimize for a particular task. If we demand the readout is higher for one `target' distribution than another, the resulting optimization just assigns those bins the maximum weight $c_\text{max} = 1$, and all other states $c_\text{min} = 0$. In the accompanying figures we demonstrate this collation, binning, and readout. 
\textit{Locating Phase Boundaries} --- In a fixed-particle number simulation, the location of phase boundaries can be approximated by the point where the variance in the maximum cluster size peaks\\

\noindent\textbf{Generating Maximum  - Entropy Distributions}: The maximum entropy distributions we present in the main text can be defined via minimum of an objective function $\mathcal{L}_k$, that specifies constraingin a distribution up moment $\mu_k = \mu_k^{\star}$:
\begin{equation*}
\mathcal{L}_k = \log P_i - \sum_{j=0}^k{\left(\mu_j^{\star}-\sum_{n=0}^{N_\text{sites}}{n^jP(n) }\right)}
\end{equation*}
where the $\mu_j^{\star}$ are the chosen target moments, and the moment up to and including $k$ are fixed to some target value. Here $\mu_0 = 1$ in all cases for normalization, and we constrain to $P_i \ge 0$. In our examples we also typically fix $\mu_1 = \frac{N_\text{sites}}{2} = 3$ to center the distributions. We numerically optimize the $\mathcal{L}$ over the variables $P_i$.

\section{Percolation Decoder}\label{sec:perc}
\noindent\textbf{Model} --- Percolation is another collective decoder. We consider a population of molecules, each carrying $n$ identical binding sites, where $n$ is drawn independently from a distribution $P(\mathbf{n})$ with support on $\{0, 1, \ldots, N_\text{S}\}$. Binding sites pair randomly in the well-mixed limit: any unoccupied site can encounter any other, and every encounter forms a bond. The resulting network of bond-linked molecules is a random graph with degree distribution $P(\mathbf{n})$.

The readout is the gel fraction $S$, defined as the fraction of molecules in the largest connected component. When $S$ is macroscopic (a finite fraction of $N$ as $N \to \infty$), the system has gelled: a network-spanning cluster has formed. This is a geometric connectivity transition that depends on the degree distribution $P(\mathbf{n})$.\\

\noindent\textbf{Locating the Percolation Transition} --- The probability generating function for the degree distribution is $G_0(x) = \sum_{n} P(n)\, x^{n}$, with mean $\mu = \langle n \rangle = G_0'(1)$ and variance $\sigma^2 = \langle n^2 \rangle  - \mu^2 = G_0''(1) + \mu(1-\mu)$. When we follow a randomly chosen bond to one of its endpoints, the number of additional bonds at that endpoint is distributed according to the excess degree distribution, with generating function $G_1(x) = G_0'(x)/G_0'(1)$. A macroscopic cluster appears when the expected number of onward connections, or branching ratio $\alpha$, exceeds one:

\begin{equation*}
\alpha \equiv G_1'(1) = \frac{\langle n(n-1) \rangle}{\langle n \rangle} > 1
\end{equation*}

which is the Molloy-Reed criterion\cite{molloy1995critical}. Rewriting in terms of distribution moments $\mu$ and $\sigma^2$, the gelation condition becomes:

\begin{equation*}
\sigma^2 > 2\mu - \mu^2
\end{equation*}

The gel point contour in the $(\mu, \sigma^2)$ plane is therefore the parabola $\sigma^2_\text{perc} = \mu(2 - \mu)$, with roots at $\mu = 0$ and $\mu = 2$ and maximum $\sigma^2 = 1$ at $\mu = 1$. For mean valency $\mu \geq 2$, gelation occurs at any variance; for $\mu < 2$, sufficient polydispersity is required.\\

\noindent\textbf{Gel Fraction} --- Above the gel point, the gel fraction can be determined self-consistently. We assume that the network contains no loops and is locally tree-like. Let $u$ be the probability that following a random bond fails to reach a system-spanning cluster. A molecule with $n$ bonds does not reach the system spanning cluster if all $n$ outward bonds also fail to connect to the spanning cluster, which occurs with probability $u^n$. Averaging this probability over the excess degree distribution gives back $G_1(u)$, defined above:

\begin{equation*}
u = \sum_n P_\text{exc}(n)  u^{n} = G_1(u)
\end{equation*}

We take the smallest non-negative root of this equation. The value $u = 1$ always solves this equation, and below the gel point it is the only root in $[0,1]$: every bond leads to a finite cluster, so no spanning network exists and $S = 0$. Above the gel point a second root $u < 1$ appears, and this is the physical solution.

Once $u$ is known, the gel fraction follows immediately. A molecule lies in a finite cluster (the sol) only if all $n$ of its own bonds lead to finite subtrees, with probability $u^{n}$. Averaging over the valency distribution $P(n)$ gives the sol fraction $G_0(u)$, so the gel fraction is its complement:

\begin{equation*}
S = 1 - G_0(u) = 1 - \sum_{i} P(i) u^{i}
\end{equation*}

At the gel point $u = 1$ and $S = 0$; above it $u < 1$ and $S > 0$. The transition is continuous: as the branching ratio $\alpha$ increases past 1, the gel fraction grows as $S \propto  (\alpha - 1)$ to leading order.

To understand the contour geometry, we notice that the key quantity controlling gelation is the branching parameter $\alpha = \langle n(n-1) \rangle$. Define the effective single-molecule response $g(n) = n(n-1)$. This function is convex,

\begin{equation*}
g'(n) = 2n - 1, \qquad g''(n) = 2 > 0
\end{equation*}

placing gelation in the collective tail-amplifying class. From the moment expansion $\langle g(n) \rangle \approx g(\mu) + \tfrac{1}{2} g''(\mu)\,\sigma^2$, the slope of a constant-output contour in the $(\mu, \sigma^2)$ plane is

\begin{equation*}
\frac{d\sigma^2}{d\mu}\bigg|_{\langle g \rangle = g} = -(2\mu - 1)
\end{equation*}

which is negative for $\mu > 1/2$, well below the gelation threshold. Constant-gel-fraction contours therefore run from upper left to lower right in the $(\mu, \sigma^2)$ plane: increasing the variance at fixed mean pushes the system deeper into the gel phase.

\noindent\textbf{Lattice simulations} -- The mean-field theory assumes that binding sites pair globally at random, neglecting spatial structure. To test whether the favorable contour geometry survives when bonds are restricted to nearest neighbors, we performed Monte Carlo simulations on a three-dimensional cubic lattice of side length $L = 30$ sites, coordination number $z = 6$. Each site carries a molecule with valency $n$ drawn from $P(n)$, and the $n$ binding stubs are assigned to $i$ randomly chosen lattice directions. A bond forms between two adjacent sites only when both have a stub directed toward the other. For each input distribution we computed the gel fraction $S$ as the fraction of molecules in the largest connected component, averaged over 15 independent realizations per distribution point. 

\pagebreak

\section{Membrane curvature Decoder}\label{sec:membrane}
\textbf{Model} --- Membrane mechanics provides a collective decoder that is fundamentally mechanical, operating through the coupling between local protein density and membrane shape. 

We consider a flat lipid membrane carrying a dilute population of curvature-generating proteins at total area fraction $\phi_\text{tot}$. Each protein belongs to one of $N_\text{S} + 1$ variants labeled by a trait $n = 0, 1, \ldots, N_\text{S}$, where $n$ encodes the number of curvature-generating domains (amphipathic helices, BAR-domain contacts, or similar structural motifs)~\cite{zimmerberg2006proteins,johnson2024protein}. The distribution over variants, $P(\mathbf{n})$, is set by the upstream encoder. Variant $n$ has spontaneous curvature coupling $c_n = c_0\, n$: molecules with $n = 0$ are curvature-inert, while those with $n = N_\text{S}$ bend the membrane most strongly.

We begin from the most general Canham–Helfrich free energy ~\cite{helfrich1973elastic} for a fluid membrane $S$ carrying a composition field $\{\phi_n\}$, before any geometric or constitutive approximation:

\begin{equation}
F_\text{mem}[S, \{\phi_i\}] = \int_S dA\;\Big[\,\sigma_\text{mem}(\{\phi_i\}) + \frac{\kappa(\{\phi_i\})}{2}\bigl(C - C_0(\{\phi_i\})\bigr)^2 + \bar\kappa(\{\phi_i\})\,K + f_\text{mix}(\{\phi_i\})\,\Big]
\end{equation}

The integral is over the induced area element $dA$, so the surface tension $\sigma$ multiplies the true membrane area rather than its projection. The geometry enters through the total curvature $C \equiv c_1 + c_2$ and Gaussian curvature $K\equiv c_1 c_2$, with $c_1, c_2$ the principal curvatures. To minimally model coupling between membrane shape and protein density, we assume that the spontaneous curvature is set by the local composition,

\begin{equation*}
C_0(\{\phi_n\}) = c_{0,\text{mem}} + \sum_{i=0}^{N_\text{S}} c_i\, \phi_i(\mathbf{r})
\end{equation*}

where $c_{0,\text{mem}}$ is the bare membrane spontaneous curvature and $c_n = c_0\,n$ the curvature generated per protein of trait $n$. The composition free energy per unit area is given by the mixing entropy:

\begin{equation*}
f_\text{mix} = \frac{k_BT}{a_p^2}\left(\sum_{i=0}^{N_\text{S}}\phi_i \ln \phi_i + (1 - \phi_\text{tot})\ln(1 - \phi_\text{tot})\right)
\end{equation*}

with $a_p$ the protein molecular footprint, and  $\phi_\text{tot} = \sum_{i=0}^{N_\text{S}} \phi_i(\mathbf{r})$. Next we introduce several approximations to the free energy: first, we neglect the Gaussian curvature and assume that tension $\sigma_\text{mem}$ and curvature moduli $\kappa$ do not explicitly depend on protein composition. In order to analyze the membrane shape, we parameterize the surface by a height field $h(\mathbf{r})$. This allows us to write the spontaneous curvature $C = \nabla^2h$, and the area element $dA = (1+\frac{1}{2}(\nabla h))^2d^2r$. Finally, we introduce the finite protein footprint through defining $C_0 = K(\mathbf{r})*H_0({r})$, where $K(\mathbf{r})$ is the footprint profile of the proteins. With these approximations we re-write the membrane free energy, suppressing coordinates:

\begin{equation*}
F_\text{mem}[h, \{\phi_i\}] = \int d^2r \left[ \frac{\kappa}{2}\bigl(\nabla^2 h - K*H_0\bigr)^2 + \frac{\sigma_\text{mem}}{2}|\nabla h|^2 + f_\text{mix}(\{\phi_i\}) \right]
\end{equation*}


\noindent\textbf{Instability Criterion} --- To analyze instabilities of $F_\text{mem}$, we first expand the curvature energy and keep terms of $\mathcal{O}(\phi)$ and lower:
\begin{align*}
\int d^2r \frac{\kappa}{2}\bigl(\bigl(\nabla^2 h\bigr)^2 -\underbrace{2 K*\nabla^2h H_0}_{2u_s H_0}+  \bigl( K*H_0\bigr)^2\bigr) \approx \int d^2r \frac{\kappa}{2}\bigl(\bigl(\nabla^2 h\bigr)^2 -2 u_s H_0\bigr)
\end{align*}
so the full membrane free energy now reads:
\begin{equation*}
F_\text{mem}[h, \{\phi_i\}] = \int d^2r \left[ \frac{\kappa}{2}\bigl(\nabla^2 h\bigr)^2 - \kappa u_sH_0(\lbrace\phi_i\rbrace) + \frac{\sigma_\text{mem}}{2}|\nabla h|^2 + f_\text{mix}(\{\phi_i\}) \right]
\end{equation*}

We then minimize $F_\text{mem}$ with respect to $\phi_n$ at fixed $h$ to obtain the curvature - induced protein densities:

\begin{equation*}
\phi_n(\mathbf{r}) = \frac{\bar\phi_n\, e^{s\, c_n\, u_s}}{1 - \bar\phi_\text{tot} + \bar\phi_\text{tot}\,\langle e^{s\, c_i\, u_s}\rangle_P}
\end{equation*}

where $s = \kappa a_p^2 / k_BT$ and $\bar\phi_n = \phi_\text{tot}\, P(\mathbf{n})$ is the mean density of variant $n$. Proteins with large trait values $n$ accumulate in regions of positive $u_s\propto\nabla^2 h$ (concave membrane), while curvature-inert proteins ($c_n = 0$) are mildly depleted from these regions and otherwise uniformly distributed. We now linearize the protein density, expanding in powers of the curvature $s c_0 u_s$:

\begin{equation*}
\delta\phi(\mathbf{r}) \equiv \phi_i(\mathbf{r}) -\bar\phi_i\approx\bar\phi_i s\left(c_i-\phi_\text{tot}\langle c\rangle_P\right)u_s(\mathbf r)\\
\end{equation*}
Finally, we Fourier transform the above equation to obtain, and insert into $F_\text{mem}[h]$ to obtain $F_\text{mem,eff}[h(\mathbf{q})]$. The protein density in Fourier space is:
\begin{equation*}
\delta\phi_i(\mathbf{q}) = -\bar\phi_i s\left(c_i-\phi_\text{tot}\langle c\rangle_P\right)q^2 \hat K(\mathbf{q})\hat h(\mathbf{q})\\
\end{equation*}
with $\hat K(\mathbf{q})=\int d^2rK(r)e^{-i\mathbf q\cdot\mathbf r}$ and $\hat h(\mathbf{q})=\int d^2rh(r)e^{-i\mathbf q\cdot\mathbf r}$ being the Fourier transforms of the protein footprint profile and  membrane height field. Plugging this into the free energy and expanding to second order in $f_\text{mix}({\phi_i})$ gives:

\begin{equation*}
F_\text{mem,eff}[h(\mathbf{q})] =\frac12\sum_q\Bigl[\kappa q^4\underbrace{\bigl(1-s\phi_\text{tot}\big(\langle c^2\rangle_P-\phi_\text{tot}\langle c\rangle_P^2\big) \hat K(q)^2}_{\kappa_\text{eff}(\mathbf{q})}\bigr)+\sigma_\text{mem} q^2\Bigr]|\hat h_q|^2    
\end{equation*}

where $\langle c^2 \rangle_P = \sum_{i=0}^{N_\text{S}} c_i^2\, P(i)$ is the second moment of the curvature coupling. The flat membrane becomes unstable when $\kappa_\text{eff}(\mathbf{q})q^4 +\sigma_\text{mem}q^2 < 0$ for any wavevector. Using $c_n = c_0\, n$, the instability criterion at long wavelengths ($q \to 0$, $\hat K(0) \approx 1$) is

\begin{equation*}
\frac{\kappa\, \phi_\text{tot}\, c_0^2}{k_BT}\, \langle n^2 \rangle_P > 1
\end{equation*}

or equivalently $\beta\, \langle n^2 \rangle_P > 1$, where $\beta = \kappa\, \phi_\text{tot}\, c_0^2 / k_BT$ captures all membrane  parameters into a single dimensionless coupling. The instability is controlled by $\langle n^2 \rangle_P = \mu^2 + \sigma^2$, the second raw moment of the trait distribution, where $\mu = \langle n \rangle$ and $\sigma^2 = \text{Var}(n)$. Contours of constant decoder response are therefore $\mu^2 + \sigma^2 = \text{const}$ which are quarter-circles in the $(\mu, \sigma^2)$ plane with slope
\begin{equation*}
\frac{d\sigma^2}{d\mu}\bigg|_{\langle n^2 \rangle = \text{const}} = -2\mu
\end{equation*}

This is negative for all $\mu > 0$, placing the membrane curvature decoder in the collective tail-amplifying class alongside phase separation and percolation. A finite $q$ mode is unstable when $\beta\langle n^2\rangle_P > \bigl(1+\sigma_\text{mem}/\kappa q^2\bigr)\hat K(\mathbf{q})^{-2}\equiv\beta_c(q_\star)$, reached at the first-unstable wavevector $q_\star$. 

\noindent\textbf{Decoder kernel and response susceptibility} --- The natural graded response is the softening parameter $R(P) = \beta\, \langle n^2 \rangle_PK(q)^2/\left(1+\sigma_\text{mem}/\kappa q^2\right)$, which measures how close the system is to the curvature instability ($R = 1, \quad \beta=\beta_c$ at the transition). Its functional derivative with respect to the trait distribution is

\begin{equation*}
\frac{\delta R}{\delta P(n)} = \frac{\beta}{\beta_c}\, n^2
\end{equation*}

This decoder kernel is convex, zero at $n = 0$ (curvature-inert proteins are invisible to the decoder), and grows quadratically, and has no inflection point or saturation. Every additional curvature-generating domain on a protein contributes more than the previous one to the instability, amplifying the high-$n$ tail of the distribution. \\

\noindent\textbf{Numerical Simulations} --- To confirm that the instability saturates at a finite amplitude and to compute the decoder response beyond the linear stability threshold, we performed numerical simulations on a two-dimensional periodic domain membrane of $L=64$.

After elimination of the protein densities (see above), the membrane dynamics reduce to a single partial differential equation for the height field $h(\mathbf{r}, t)$,

\begin{align*}
\frac{\partial h}{\partial t} = -\Gamma\, \frac{\delta F_\text{mem,eff}}{\delta h}
\end{align*}

where $\Gamma$ is a mobility and $F_\text{mem,eff}[h]$ includes the bare bending energy, tension, and the protein-mediated coupling with Flory-Huggins mixing energy. The protein contribution involves the moment-generating function of $P$ evaluated at the local smoothed curvature, divided by the Flory-Huggins partition function $Z(\mathbf{r})=1 - \phi_\text{tot} + \phi_\text{tot}\langle e^{s\, c_i\, u_s} \rangle_P$ that enforces close-packing. We set the protein footprint profile $\hat K(\mathbf{q}) = e^{-q^2 a_p^2/4}$ with $a_p = 3$. 

This equation is integrated using a semi-implicit pseudo-spectral scheme. The bare bending and tension terms ($\kappa q^4 + \sigma q^2$) are treated implicitly in Fourier space, eliminating the stiff high-frequency stability constraint. The nonlinear protein coupling is evaluated in real space at each time step and transformed to Fourier space for the update. Each step requires one forward and one inverse fast Fourier transform of the height field, plus pointwise evaluation of the protein equilibrium at each grid point (a sum over $n + 1$ species). The scheme is run as gradient descent until the RMS curvature converges, yielding the equilibrium membrane configuration for each input distribution $P(n)$
\bibliography{refs}